# Potassium Isotope Compositions of Carbonaceous and Ordinary Chondrites: Implications on the Origin of Volatile Depletion in the Early Solar System

Hannah Bloom[1], Katharina Lodders[1], Heng Chen[1, 2], Chen Zhao[1, 3], Zhen Tian[1], Piers Koefoed[1], Mária K. Pető[4], Yun Jiang[5, 6], and Kun Wang (王昆)[1]*

[1]Department of Earth and Planetary Sciences and McDonnell Center for the Space Sciences, Washington University in St. Louis, One Brookings Drive, St. Louis, MO 63130, USA

[2]Lamont-Doherty Earth Observatory, Columbia University, Palisades, NY 10964, USA

[3]Faculty of Earth Sciences, China University of Geosciences, Wuhan, Hubei 430074, China

[4]Konkoly Observatory, Research Center for Astronomy and Earth Sciences, H-1121 Budapest, Hungary

[5]CAS Key Laboratory of Planetary Sciences, Purple Mountain Observatory, Chinese Academy of Sciences, Nanjing 210008, China

[6]CAS Center for Excellence in Comparative Planetology, China

*Corresponding author email: wangkun@wustl.edu

**How to cite:**

**ABSTRACT**

Among solar system materials there are variable degrees of depletion in moderately volatile elements (MVEs, such as Na, K, Rb, Cu, and Zn) relative to the proto-solar composition. Whether these depletions are due to nebular and/or parent-body (asteroidal or planetary) processes is still under debate. In order to help decipher the MVE abundances in early solar system materials, we conducted a systematic study of high-precision K stable isotopic compositions of a suite of whole-rock samples of well-characterized carbonaceous and ordinary chondrites. We analyzed 16 carbonaceous chondrites (CM1-2, CO3, CV3, CR2, CK4-5 and CH3) and 28 ordinary chondrites covering petrological types 3 to 6 and chemical groups H, L, and LL. We observed significant K isotope ($\delta^{41}$K) variations (−1.54 to 0.70 ‰) among the carbonaceous and ordinary chondrites. In general, the two major chondrite groups are distinct: The K isotope compositions of carbonaceous chondrites are largely higher than the Bulk Silicate Earth (BSE) value, whereas ordinary chondrites show K isotope compositions that are typically lower than the BSE value. Neither carbonaceous nor ordinary chondrites show clear/resolvable correlations between K isotopes and chemical groups, petrological types, shock levels, cosmic-ray exposure ages, fall/find occurrence, or terrestrial weathering. Importantly, the lack of a clear trend between K isotopes and K content among chondrites indicates that the K isotope fractionations were decoupled from the relative elemental K depletions, which is inconsistent with a single-stage partial vaporization or condensation process to account for these MVE depletion patterns among chondrites. The range of K isotope variations in the carbonaceous chondrites in this study is consistent with a four-component (chondrule, refractory inclusion, matrix and water) mixing model that is able to explain the bulk elemental and isotopic compositions of the main carbonaceous chondrite groups, but requires a fractionation in K isotopic compositions in chondrules. We propose that the major control of the isotopic compositions of group averages is condensation and/or vaporization in pre-accretional (nebular) environments that is preserved in the compositional variation of chondrules. Parent-body processes, such as aqueous alteration, thermal metamorphism, and metasomatism, can mobilize K and affect the K isotopes in individual samples. In the case of the ordinary chondrites, the full range of K isotopic variations can only be explained by the combined effects of the size and relative abundances of chondrules, parent-body aqueous and thermal alteration, and possible sampling bias.

# 1. INTRODUCTION

Compared to the bulk composition of our solar system, represented by the solar photosphere and CI chondrites (Palme et al., 2014), other chondritic meteorites and in particular differentiated planetary bodies of the inner solar system such as the Earth, Moon, Mars, and asteroid 4 Vesta are depleted in moderately volatile elements (*e.g.*, Davis 2006; Palme et al. 1988) (see **Figure 1**). Conventionally, the volatility of an element is defined by its 50 % nebular condensation temperature (*Tc*) calculated with the initial solar nebula composition at $10^{-4}$ bars (*e.g.*, Grossman and Larimer 1974; Lodders 2003). The "moderately volatile elements" (MVEs) are elements with 50 % *Tc* between 650 K and 1250 K, such as Na, K, Rb, Cu, and Zn. The MVEs of undifferentiated bodies (*i.e.*, chondrites) are generally depleted within one order of magnitude compared to the CI chondrites, while those of differentiated planetary bodies such as the Moon and Vesta can be depleted by as much as two orders of magnitude (see **Figure 1**). These MVE depletions across the solar system are an intrinsic property of the overall solar system chemical structure.

Two scenarios have been suggested to explain these observations: One ascribing the fractionations to processes occurring in the solar nebula environment and the other to planetesimal/planet formation. Nebular MVE depletion mechanisms include:

1) initial incomplete condensation from the solar nebula (Wasson and Chou, 1974; Wai and Wasson, 1977; Humayun and Clayton, 1995; Cassen, 1996; Ciesla, 2008),

2) partial evaporation of interstellar dusts prior to incorporation into the solar nebula (Yin, 2005) or in the solar nebula (Huss et al., 2003; Huss, 2004);

3) mixing of chondrite components formed in distinct volatile-rich and volatile-

poor reservoirs of the early Solar System (e.g., Alexander et al. 2001; Anders 1964; Shu et al. 1997);

Processes depleting MVEs during planetesimal/planet formation include:

4) accretional volatile loss (Ringwood, 1966; Albarède, 2009; Hin et al., 2017);

5) giant impacts (Paniello et al., 2012; Wang and Jacobsen, 2016b);

6) magma ocean degassing (Day and Moynier, 2014; Kato et al., 2015);

7) extraction into steam atmospheres and atmospheric loss (Fegley et al., 2016; Young et al., 2019).

If the volatile depletions occurred mainly in nebular environments, and planetary bodies simply inherited these depletions from their building-blocks (possibly some material akin to chondrites and their components), we should expect to observe at least some trends of volatile element depletions among undifferentiated, primitive meteorites. Indeed, such trends are seen for the different types of carbonaceous chondrites (CCs). However, the degree of depletion among primitive meteorites are much less than those observed in differentiated planetary materials. Compared to other chondrites, CI chondrites are not volatile-depleted, and their compositions are similar to that of the Sun (except for the highly volatile elements H, C, N, O, and noble gases). The abundances of volatile elements in other CCs decrease in the sequence of CI>CM>CO=CV>CR>CK (Lodders and Fegley, 1998; Lodders, 2003) (see MVEs in **Figure 1**). Unlike the patterns of volatile depletions among differentiated planetary materials, the volatile depletions in the CCs are plausibly explained by fractional condensation (or evaporation) in the nebular environment. Among the ordinary chondrites (OCs) the relative MVE abundances are quite similar but clearly different from the trends in the CC groups (see **Figure 1**). Ordinary chondrites were subjected to various degrees of thermal alteration in their parent bodies, which caused larger heterogeneities in the meteorite samples for highly volatile elements (50 % $Tc$ <650 K; e.g., Cd, In, Bi and Tl), while the abundances of the MVEs (except for siderophile MVEs) do not vary dramatically among the groups and petrological types of the OCs (Wasson and Chou, 1974).

The isotopic compositions of the MVEs are promising tools for determining and distinguishing between the various processes proposed for producing the MVE fractionations (*e.g.*, Humayun and Clayton 1995). Incomplete condensation or partial evaporation under different P-T conditions would generate different magnitudes of either equilibrium or kinetic isotopic fractionations (*e.g.*, Richter et al. 2009). For example, upon heating dust at low ambient vapor pressures $P_{vap}$ (<<$P_{sat}$; saturation vapor pressure), net evaporation and kinetic isotope fractionation become dominant. In contrast, when $P_{vap}$ =$P_{sat}$, the system reaches equilibrium and only (small) equilibrium isotopic fractionation occurs. Thus, studying the isotopic compositions of the MVEs in CCs and OCs could constrain the conditions (*e.g.*, ambient vapor pressure) where the MVEs were depleted.

Potassium is a moderately volatile element with a 50 % $Tc$ of ~1006 K (Lodders, 2003). The isotopes of K have been studied as a potential tracer of volatile depletion as early as 1995 (Humayun and Clayton, 1995). At that time, no variations between differentiated and undifferentiated meteorites were observed due to the relatively large analytical uncertainties using Secondary Ion Mass Spectrometry (SIMS). Since 2016, high-precision K isotope analytical methods using Multi-Collector Inductively-Coupled-Plasma Mass-Spectrometry (MC-ICP-MS) (Wang and Jacobsen, 2016a; Li et al., 2016; Morgan et al., 2018; Hu et al., 2018) revealed resolvable variations in K isotopic compositions between different planetary materials (*e.g.*, Earth and Moon) that were subsequently used to better understand the accretion mechanism of different planetary bodies.

Only a few high-precision K isotope data have been reported for chondrites since 2016: data are available for one CC (Orgueil, CI1) and three equilibrated OCs (Guareña, H6; Bruderheim and Peace River, L6) (Wang and Jacobsen, 2016b). Previous isotope studies of other moderately volatile elements such as Zn and Rb (Pringle et al., 2017; Pringle and Moynier, 2017) report relatively large isotopic fractionations among different CC and OC groups.

Here we present the first comprehensive dataset for high precision K isotopes for 16 CCs and 28 OCs, which include most chemical and petrologic types of chondrites. This study will help compare chondritic materials to differentiated planetary materials, such as Earth, Moon, Mars and Vesta, by exploring and comparing the ranges of K isotopic fractionations.

## 2. SAMPLES and METHOD

The sample names, chemical groups, petrological types, shock levels, and weathering grades of all the chondrites studied here are listed in **Table 1** and **2**. The acronyms in meteorites' names represent the places where they were found (*e.g.*, ALH, EET, GRA, GRO, LAP, LAR, LON, MAC, MET, MIL, PCA, PRA, PRE, QUE, RBT, and RKP represent Allan Hills, Elephant Moraine, Graves Nunataks, Grosvenor Mountains, LaPaz Icefield, Larkman Nunatak, Lonewolf Nunatak, MacAlpine Hills, Meteorite Hills, Miller Range, Pecora Escarpment, Mount Pratt, Mount Prestrud, Queen Alexandra, Roberts Massif, and Reckling Peak, respectively). Our sample set covers all major chemical and petrological types of CC and OC. Eight of our samples are falls and 36 samples are finds from Antarctica. To minimize the potential effects from terrestrial weathering, we include only Antarctic finds that had been exposed to little or no weathering (classified as weathering grade "A" seen in **Table 1** and **2**). Previous studies have shown that the Antarctic meteorites are well preserved and have little terrestrial contamination (*e.g.*, Crozaz and Wadhwa 2001). Tian et al. (2019) demonstrated that the K isotopes of Antarctic meteorites are indistinguishable from those of meteorite falls of the same type, in contrast to hot desert (*e.g.*, NWA Northwest Africa) meteorites, which are significantly contaminated by terrestrial K or/and altered due to terrestrial weathering.

Details of our analytical method can be found in Chen et al. (2019). For all samples, chips (free of fusion crust) were selected and finely ground using an agate pestle and mortar. About 100-200 mg of the well-mixed powders were dissolved for K isotope analysis (see **Table 1** and **2** for the mass dissolved for each sample). All samples were digested using Parr high-pressure digestion vessels at 150 °C in a Quincy Lab Model 10 lab oven for two days through a two-step protocol (first in concentrated HF/HNO$_3$ mixture and then in aqua regia). Fully digested samples were dried down and refluxed with 0.7 N HNO$_3$. Then samples were loaded into chromatography columns (ID=1.5 cm; filled with 17 mL Bio-Rad AG50W-X8 100-200 mesh cation exchange resin). Potassium was eluted and collected with 0.7 N HNO$_3$. To further purify K from matrix elements, we repeated the same procedure a second time using a smaller column (ID=0.5 cm; filled with 2.4 mL resin). The yield of the procedure is > 99 %. The total-procedure blank is 0.26 ± 0.15 µg (2SD; n=7; Chen et al. 2019) and is negligible (<1 %) compared to the amount of K in the samples.

The concentrations of K and matrix elements after columns were measured with a Thermo Scientific iCapQ quadrupole ICP-MS. The K isotopic compositions were measured with a Thermo Scientific Neptune Plus MC-ICP-MS using the sample-standard bracketing (SSB) method. All K isotopic data is expressed in the delta notation, where $\delta^{41}K=[(^{41}K/^{39}K)_{sample}/(^{41}K/^{39}K)_{NIST\ SRM\ 3141a}-1]\times1000$. To monitor the data quality of each analytical session (24 hours), we measured a geostandard (BHVO-2) along with the chondrite samples. The typical within-session reproducibility is ~0.05 ‰ (95 % c.i.; confidence interval). The long-term (~20 months) reproducibility of this method has been evaluated as 0.11 ‰ (Chen et al. 2019), which is our most conservatively estimated analytical error.

# 3. RESULTS

Results for the CCs are given in **Table 1** and the OCs in **Table 2**, with data for the geostandard in each analytical campaign given in the respective tables. All individual sample errors are shown as 95 % c.i. while all sample populations are shown as 2 standard deviation (2SD). Our determination of the K isotopic composition ($\delta^{41}$K) of the basaltic geostandard BHVO-2 ($-0.47 \pm 0.03$ ‰ and $-0.49 \pm 0.03$ ‰; see **Table 1** and **2**) agrees well with published data (ca. $-0.5 \pm 0.05$ ‰; **Hu et al. 2018; Li et al. 2016; Morgan et al. 2018; Wang and Jacobsen 2016b**), which gives us confidence in the accuracy of our measurements.

## 3.1 Potassium isotopic compositions of the carbonaceous chondrites

**Figure 2a** illustrates the significant K isotopic fractionations in the different chemical groups of the CCs. The range of $\delta^{41}$K for all 17 CCs span from $-1.54 \pm 0.03$ ‰ to $0.70 \pm 0.02$ ‰. Excluding three extreme cases (MIL 11213 (CO3), LAR 12002 (CV3) and GRA 06100 (CR2)) the other 14 CCs span a much more restricted range of $\delta^{41}$K from $-0.64 \pm 0.03$ ‰ to $-0.06 \pm 0.04$ ‰, with an average (simple arithmetic mean) of $-0.29 \pm 0.34$ ‰ (2SD).

The CI chondrite $\delta^{41}$K value of $-0.53 \pm 0.11$ ‰ from **Wang and Jacobsen (2016a)** is within the uncertainty of the Bulk Silicate Earth (BSE in **Figure 2a**). The four CM meteorites have a $\delta^{41}$K range from $-0.39 \pm 0.03$ ‰ to $-0.09 \pm 0.03$ ‰, with a mean of $-0.25 \pm 0.26$ ‰ (2SD). The range for the four CO meteorites is $-0.35 \pm 0.07$ ‰ $\leq \delta^{41}$K $\leq 0.36 \pm 0.05$ ‰, averaging $-0.09 \pm 0.58$ ‰ (2SD). One CV chondrite (LAR 12002) has the lowest $\delta^{41}$K value of all samples analyzed in this study; the low value was confirmed with a second fragment from this meteorite (−1.45 ± 0.04 ‰ and −1.54 ± 0.03 ‰). The other two CV meteorites have a $\delta^{41}$K range from −0.53 ± 0.03 ‰ to −0.30 ± 0.02 ‰, averaging at −0.39 ± 0.25 ‰ (2SD). Only two fragments of a single CR meteorite (GRA 06100) were analyzed. Both fragments show high $\delta^{41}$K values (0.70 ± 0.02 ‰ and 0.16 ± 0.03 ‰) with one of the two fragments showing the highest $\delta^{41}$K value (0.70 ± 0.02 ‰) of all the samples analyzed in this study. The four CK meteorites have a $\delta^{41}$K range from −0.36 ± 0.03 ‰ to −0.06 ± 0.04 ‰, averaging at −0.15 ± 0.25 ‰ (2SD). The only CH meteorite in our study yields a $\delta^{41}$K of −0.64 ± 0.03 ‰, slightly lighter than the BSE value. Of the CC samples analyzed here, the majority have *heavier* K isotopic compositions than that of the CI chondrites (−0.53 ± 0.11 ‰; Wang and Jacobsen 2016a) and the BSE (−0.48 ± 0.03 ‰; Wang and Jacobsen 2016b).

## 3.2 Potassium isotopic compositions of the ordinary chondrites

Figure 2b shows large variations in the K isotopic compositions ($\delta^{41}$K varying from −1.47 ± 0.06 ‰ to 0.07 ± 0.03 ‰) of the OCs that were measured here. The ranges of $\delta^{41}$K of the three OC groups (H, L, LL) overlap; however, on average the H group has slightly heavier compositions due to two H chondrites showing $\delta^{41}$K ≥ 0 (RKPA 78004 and PRE 95400). The H group has a $\delta^{41}$K range from −1.28 ± 0.05 ‰ to 0.07 ± 0.03 ‰ with an average $\delta^{41}$K of −0.70 ± 0.81 ‰ (2SD). The L group range is −1.14 ± 0.03 ‰ ≤ $\delta^{41}$K ≤ −0.54 ± 0.10 ‰ with an average $\delta^{41}$K of −0.82 ± 0.32 ‰ (2SD), and the LL group range is −1.47 ± 0.06 ‰ ≤ $\delta^{41}$K ≤ −0.44 ± 0.05 ‰ with an average $\delta^{41}$K of −0.95 ± 0.61 ‰ (2SD). Three OCs (one H6 and two L6) were analyzed previously (Wang and Jacobsen, 2016b) and they fall into the range of the OCs measured in this study. Our measurements show no systematic variations of K isotope compositions as a function of chemical group. The majority of the OC samples analyzed here have *lighter* K isotopic compositions than that of the CI chondrites (−0.53 ± 0.11 ‰; Wang and Jacobsen 2016a) and BSE (−0.48 ± 0.03 ‰; Wang and Jacobsen 2016b).

### 3.3 Potassium isotopic heterogeneity in carbonaceous and ordinary chondrites samples

To examine the K isotopic heterogeneity within the same meteorite, for 16 different meteorites we conducted measurements two or three times on different fragments (~0.1 g each fragment; labeled as #1, #2, #3 in Table 1 and 2) of the same samples (~0.5-1 g each sample). Individual fragments were powdered, dissolved, purified and analyzed independently. Within our long-term reproducibility of 0.11 ‰ (Chen et al., 2019), different fragments of 9 out of 16 meteorites yield indistinguishable K isotope compositions. The different chips of the other 7 meteorites do not agree with each other within analytical uncertainty. The $\delta^{41}K$ differences between different fragments of these 7 samples vary by 0.14 ‰ (L3.6 GRO 06054), 0.18 ‰ (LL5 RBT 04127), 0.19 ‰ (CV3 Allende), 0.27 ‰ (H4 Weston), 0.27 ‰ (L5 Homestead), 0.34 ‰ (CO3 MIL 11213), and 0.54 ‰ (CR2, GRA 06100). These compositional differences exceed analytical uncertainties and likely reflect the K isotope heterogeneity at the 0.1-gram-scale within the samples. Note that the observed heterogeneity shows no relation to the sample type of CC and OC (e.g., CV, CO, CR, H, L and LL). The causes of this K isotopic heterogeneity will be discussed in Section 4.2 and Section 4.3.

## 4. DISCUSSION

### 4.1 Effects of terrestrial weathering on potassium isotopes of chondrites.

We chose to study both falls and finds to examine whether terrestrial weathering would affect the K isotope compositions of chondrites. Chondritic meteorites only contain 200-970 ppm K (Lodders and Fegley, 1998), which is significantly lower than the K abundance on the surface of the Earth (*e.g.*, continental crust ~23244 ppm; Rudnick and Gao 2014). Terrestrial weathering and contamination could potentially add terrestrial K to meteorite finds. "Hot desert" meteorite finds have been shown to have K addition compared to meteorite falls and Antarctic finds (Bischoff et al., 1998; Zipfel et al., 2000). In addition, K salts in chondrites (*e.g.*, sylvite and halite; Rubin et al. 2002; Zolensky et al. 1999) are very soluble and could possibly be leached out during terrestrial weathering by rain or surface water. As a result, the K isotopic compositions of meteorite finds can be potentially altered.

Ideally, only samples recovered quickly from observed meteorite falls and well-curated samples should be analyzed to circumvent potential terrestrial contamination and/or alteration. Nevertheless, Antarctic meteorites finds have generally been subjected to much less terrestrial weathering than "hot desert" meteorites (*e.g.*, Crozaz and Wadhwa 2001). Furthermore, Tian et al. (2019) showed that the K isotopic compositions of eucrite meteorites from hot desert regions (*e.g.*, NWA) have significantly higher K concentrations than eucrite falls and eucrite meteorites from Antarctica. The K isotopic compositions of NWA eucrite meteorites were also shown to be similar to those of terrestrial rocks and different from those of eucrite falls (Tian et al., 2019). In contrast, all eucrite meteorites from Antarctica have indistinguishable K isotopic compositions compared to eucrite falls. In general, eucrite meteorites contain a lower abundance of K (25 to 328 ppm; Lodders and Fegley 1998) when compared to chondritic meteorites (82 to 796 ppm; **Table 1**). Thus, chondritic meteorites from Antarctica are even less likely to be affected by terrestrial weathering when compared to eucrite meteorites from Antarctica. As such, we primarily studied Antarctic finds combined with available falls and completely avoided meteorites from "hot deserts".

As shown in **Table 1** and **Figure 3**, there is no significant difference in the K isotope ranges between falls and Antarctic finds of the same chemical group. For example, the average $\delta^{41}K$ values for falls and Antarctic finds are $-0.92 \pm 0.50$ ‰ (2SD) vs. $-0.49 \pm 0.86$ ‰ (2SD) for the H chemical group; $-0.78 \pm 0.44$ ‰ (2SD) vs. $-0.84 \pm 0.19$ ‰ (2SD) in the L chemical group; and $-0.94 \pm 0.64$ ‰ (2SD) vs. $-1.04 \pm 0.08$ ‰ (2SD) for LL chemical group respectively. For the CV group, the averages (excluding the outlier LAR 12002) are $-0.44 \pm 0.27$ ‰ vs. $-0.30 \pm 0.02$ ‰, respectively. All Antarctic meteorites have been assigned a letter to indicate their weathering grades. Letters A, B or C represents "minor", "moderate" or "severe" rustiness, respectively. An additional letter "e" denotes the presence of visible evaporite minerals. **Figure 3** plots all the samples, K isotopic compositions vs. their weathering grades, and shows no correlation in any of the chemical groups. Therefore, we concluded that the majority of Antarctic meteorites (except the outlier LAR 12002) studied here were not measurably contaminated or altered by terrestrial weathering. Terrestrial weathering is a potential problem for acquiring the pristine K isotopic compositions of meteorite finds, however this is mostly restricted to meteorites from hot-desert regions. The K isotope variations we observe here are not likely due to terrestrial weathering, contamination, or alteration. As discussed in the **Online Supplementary Material**, these variations can not be attributed to the cosmic-ray radiation effect either. Therefore, these K isotope variations are more likely due to nebular or parent-body processes.

## 4.2 Potassium isotope fractionation of carbonaceous chondrites

For the CCs, there is a long-recognized trend of volatile depletion among the different chemical groups (Wasson and Chou, 1974; Wai and Wasson, 1977). As discussed in the **Introduction** and shown in **Figure 1**, this trend has the same monotonic depletion pattern starting from CI to CM, CO, CV, CR and CK for all MVEs such as Na, K, Rb, Cu, and Zn. The degrees of MVE depletion are mostly controlled by nebular volatilities (*e.g.*, Lodders 2003) rather than by their chemical affinities (*i.e.*, lithophile, siderophile, and chalcophile). Thus, this volatile depletion trend exhibited in the CCs has been attributed to solar nebular processes through either incomplete condensation (Wasson and Chou, 1974; Wai and Wasson, 1977; Humayun and Clayton, 1995; Cassen, 1996; Ciesla, 2008), or partial vaporization (Huss et al., 2003; Huss, 2004);

Previously, K isotopes have been used to distinguish between these two processes (Humayun and Clayton, 1995). As explained by Humayun and Clayton (1995), if the observed trend was caused by incomplete condensation (assuming near equilibrium conditions), we would expect no measurable K isotope fractionation between volatile-rich CI and volatile-depleted CC groups. In contrast, partial vaporization involving Rayleigh distillation (a kinetic process) would enrich heavy K isotopes in depleted CC groups, for example, ~25 ‰ enrichment for CV and ~20 ‰ for CM relative to CI (Humayun and Clayton, 1995). The five CCs (one CI, Orgueil; one CM, Murchison; and three CV, Allende, Vigarano, and Leoville) previously studied by Humayun and Clayton (1995) showed no significant K isotope fractionation, thus they concluded that partial vaporization is inconsistent with the K isotopes and incomplete condensation from a hot solar nebula better accounts for the volatile depletion observed among the CC groups.

In this study, we analyzed 16 CCs and observed a ~2 ‰ variation among the CC samples with a typical 0.05 ‰ analytical uncertainty (see **Figure 2**). These new results agree with Humayun and Clayton (1995)'s study of the variation of K isotopes among the CCs. In their study, the difference in the K isotope composition between the highest (1.5 ± 0.5 ‰; CV3 Leoville) and lowest values (−0.5 ± 0.7 ‰; CV3 Allende) is 2.0 ± 0.9 ‰. However, considering the error bars of 0.5-0.7 ‰, their data was interpreted as isotopically homogenous among all chondrites. Here, we confirm the range of K isotopic variation first observed by Humayun and Clayton (1995), but suggest that these variations among the CCs are indeed resolvable. Our data also agree with Humayun and Clayton (1995) in that there is no well-defined trend of heavy K isotope enrichments with increasing degrees of elemental depletion (from CI to CM, CO, CV, CR and CK; see **Figure 2**). This lack of a clear trend in K isotopes among the CCs supports Humayun and Clayton (1995)'s conclusion that a single-stage partial vaporization process cannot generate both the K element and K isotopic variations exhibited among the CC groups. Alternatively, Young (2000) argued that diffusion-limited evaporation (e.g., evaporation of solid dust) would also suppress the K isotopic fractionation. In addition, evaporation experiments at higher pressures (*e.g.*, 1 bar) also show much smaller K isotopic fractionation compared to evaporation experiments under low-pressure (~$10^{-4}$ bar) nebular conditions (Yu et al., 2003; Richter et al., 2011). Thus, the suggestion that partial vaporization resulted in the volatile depletion trend exhibited in the CCs cannot be entirely ruled out.

Although there is no trend or correlation between K isotope compositions and K concentrations (or K/U, K/Al ratios) among CCs (this is also true for OCs), we note that there seems to be a rough inverse correlation when all chondrite groups (CCs + OCs) are considered together (see **Figure 4**). This rough inverse correlation clearly extends from the CCs to the OCs, and the correlation coefficients ($r^2$) are 0.453 (K/U), 0.586 (K/Al), and 0.615 ([K]). One could argue that the correlations with the K/U and K/Al ratios are due to changes in the abundances of the refractories (U and Al) in the chondrites. However, the absolute concentrations of K also show an inverse correlation with K isotopic composition (**Figure 4c**). Because our findings particularly highlight the larger $\delta^{41}K$ variations among meteorites within the same chemical groups rather than the variations between the average compositions of the different chondrite groups (see **Figure 2**), the correlation is largely driven by the clear differences in the isotopic compositions of the CCs (on average heavier than BSE) compared to the OCs (on average lighter than BSE). This K isotope dichotomy between the CCs and the OCs will be discussed in detail in **Section 4.5**.

The members of each CC group exhibit a large range of K isotopic compositions (~1 ‰; for example, CM, CO, and CV see **Figure 2**) regardless of their degree of volatile depletion relative to CI. These intra-group variations are a significant reason for why a clear, well-defined trend is not discernable. Since K is fluid-mobile and moderately volatile, it is susceptible to modification in the parent bodies and heterogeneity could be introduced by the redistribution of K. Thus, the range in K isotope fractionations within a meteorite group observed here might be the result of parent-body processes such as aqueous alteration, thermal metamorphism, and impact vaporization that overprinted any well-defined volatile element fractionations stemming from nebular processes recorded in the chondrite components that accreted to the different chondrite parent bodies.

*Aqueous alteration and thermal metamorphism*: Petrological types (1-6) of chondrites are commonly used as a general indicator for the degree of aqueous alteration and thermal metamorphism, where Type 3 represents the most primitive (unequilibrated) type. From Type 2 to 1, the degree of aqueous alteration increases; while from Type 3.0 to 6, the degree of thermal metamorphism increases. **Figure 5a** plots the K isotopes of the bulk CCs versus their petrological types. We find that the average K isotope compositions of petrologic types are not significantly different from each other, which is consistent with a closed-system and isochemical modification (Kerridge et al., 1979; McSween, 1979; Bland et al., 2009). As shown in **Figure 5a**, Type 3 CCs exhibit the largest variations in K isotopic compositions while, with increasing aqueous alteration (1-2) and thermal metamorphism (4-5), the K isotopes of different meteorites in the same chemical group become less dispersed. It appears both aqueous alteration and thermal metamorphism in the parent bodies have affected and likely homogenized the initial K isotopic compositions inherited from the components in the solar nebula that eventually accreted into the meteorite parent bodies (as shown in petrologic Type 3). However, this observation needs to be further confirmed with a larger dataset for each chemical group since currently the number of measured samples is limited (due to sample availability).

*Impact vaporization*: Another possible source of isotopic fractionation is impact vaporization (e.g., Hin et al. 2017; Housley 1979; Moynier et al. 2011). Impact generated melting and vaporization have been invoked to explain the losses of the more volatile MVEs, Cu and Zn, and associated isotopic fractionations (Moynier et al., 2009; Moynier et al., 2010a; Moynier et al., 2010b). The shock degrees of meteorites are indicative of the strength of the impact events on the parent body asteroids (see **Table 1**). Although the evaluation of the shock levels experienced by the CCs used in this study is very limited, we can examine the shock effect on K isotopes in the OCs (see **Section 4.3**). There is no correlation between the shock degrees and K isotopes for samples from the same chemical groups and petrological types (see **Table 1**). In general, the heterogeneity of the K isotopic compositions within a chondrite group does not seem to be the result of different degrees of impact vaporization. Nevertheless, we need to note that GRA 06100 does show evidence of shock metamorphism and subsequent shock-induced hydrothermal alteration (Abreu and Bullock, 2013). As shown in **Table 1**, the two chips of GRA 06100 show significantly different K isotope compositions (0.16 ± 0.03 ‰ vs. 0.70 ± 0.02 ‰). Although the estimated peak temperature of shock metamorphism (~600 °C; Abreu and Bullock, 2013) is too low to be able to drive off the MVEs from the CR parent body, the shock-induced hydrothermal alteration could have fractionated K isotopes (Parendo et al., 2017; Li et al., 2019), which may explain the heterogeneity seen in the K isotopes of the two chips of GRA 06100.

It seems that parent body processing cannot explain the major K isotopic variations observed here, *e.g.*, Type 3 CCs show the largest K isotopic variations (see **Figure 2**). While parent-body processes may have equilibrated the K isotopes and reduced the variability (see **Figure 5a**), pre-accretionary processes likely played a major role in shaping the K isotope fractionation observed here. Recently, Alexander (2019) proposed a quantitative model that has successfully explained the elemental and isotopic fractionations among CCs. Here we apply this model to K isotopes.

*Four-component-mixing model for K isotopes*: Alexander (2019) proposed that variable mixing of the same four components can reproduce the bulk elemental compositions and isotopic anomalies of the six CC groups (CI, CM, CO, CV, CR and CK). The four components are chondrules, refractory inclusions, matrix, and water. The model predicts the mass fractions of each component for the six CC groups (Table 5 in Alexander 2019) and estimates the end-member composition of each component (Table 6 in Alexander 2019). The model data can reconstruct the bulk K elemental compositions for the six CC groups (see **Figure 6a**). Here, we use our results to assign K isotopic compositions to each end-member component and to reconstruct the K isotopic compositions of the six CC groups (see **Figure 6b**). We take the K isotopic composition of a CI chondrite (−0.53 ± 0.11 ‰; Wang and Jacobsen 2016a) as the composition of the matrix component. For the refractory inclusions and water components, there is no K isotopic composition assigned since these two components do not (nominally) contain K (see Table 6 in Alexander 2019). For the K isotopic composition of the chondrule component we have estimated a range of −0.282 ‰ to 2.076 ‰ based on both theoretical and empirical considerations as explained below.

The measurements of K isotope compositions of individual chondrules have been conducted *in situ* using Secondary Ion Mass Spectrometry (SIMS) on Type 3 chondrites and show large variability (Alexander et al., 2000; Alexander and Grossman, 2005). Due to the isobar interferences and the matrix effect associated with the SIMS technique, the large variability was (at least partially) attributed to analytical artifacts (Alexander et al., 2000; Alexander and Grossman, 2005). However, chondrules from Type 3.0 chondrites (with little parent-body alteration) show a K depletion that is correlated with chondrule mass/size (Hewins et al., 1997) with larger chondrules showing higher depletions in K. This was attributed to more K volatilization during chondrule melting. We thus predict that larger chondrules are enriched in heavier K isotopes because volatile loss would enrich heavier K isotopes in the residual melt, as demonstrated by laboratory experiments (Yu et al., 2003; Richter et al., 2011). This isotope fractionation mechanism of evaporation during chondrule formation has been previously used to interpret the Fe isotope fractionation among individual chondrules from CM, CV and ordinary chondrites (Mullane et al., 2005; Needham et al., 2009; Hezel et al., 2010; Hezel et al., 2018b). With the evidence in both observation and experiments discussed above, we would expect the chondrule component in the mixing model to be enriched in heavier K isotopes during chondrule formation events. As shown in the vaporization experiments (Yu et al., 2003; Richter et al., 2011), K isotope fractionation (in low pressure and dust density nebular environments) is very sensitive to the degree of K loss. A few % loss would induce a large (a few ‰) K isotope fractionation. Following the Rayleigh distillation law and using the ideal Rayleigh fractionation factor (square root of 41/39) in vacuum (*e.g.*, a typical nebular pressure ~ $10^{-4}$ bar), we estimated a range of K isotope compositions for the possible chondrule component as from −0.282 ‰ to 2.076 ‰ depending on the fraction loss of K (from 1 % to 10 %; labeled in **Figure 6b**) during chondrule formation.

The maximum 10 % loss of K during chondrule formation is constrained by the Na content in chondrules (Alexander et al., 2008). The estimated range of K isotopes among chondrules in this mixing model is consistent with the range (a few ‰) measured in individual chondrules (Alexander et al., 2000; Alexander and Grossman, 2005).

We want to note that the loss of volatiles *during* the chondrule formation is insignificant (typically within a few %; up to 10 %), which is consistent with Alexander (2019)'s model. Alexander (2019) proposed that the chondrule precursor had already lost MVEs *prior to* chondrule formation (melting and evaporation) as shown in the Fig. 14 in Alexander (2019) most likely through incomplete condensation (Wai and Wasson 1977; Wasson and Chou 1974), where no K isotopic fractionation is expected (Humayun and Clayton, 1995). The MVE depletion likely predated the chondrule formation events based on two constraints. 1) The abundances of MVEs in the chondrule component correlate well with their predicted 50 % condensation temperatures (Wai and Wasson 1977; Wasson and Chou 1974). Therefore, volitalization during chondrule formation cannot be a dominant process for the MVE depletion as it would lead to a poor correlation between MVEs contents and their predicted 50 % *condensation* temperatures. 2) The K isotopic compositions of chondrules would otherwise show much more extreme variaibility (up to 100 ‰; Humayun and Clayton (1995)), which has not been observed (Alexander et al., 2000; Alexander and Grossman, 2005).

As explained above, we have applied the four-component mixing model from Alexander (2019) to K isotopes. **Figure 6b** shows that the model result in general agrees well with the measurements. For example, as shown in **Figure 6b**, CR chondrites show the heaviest K isotope compositions. Hezel et al. (2018a) recently compiled the median K elemental compositions of bulk chondrules from each chondrite group, and CR chondrules have the lowest K concentration, which is consistent with our model prediction (*i.e.*, more K depletion results in higher K isotopic fractionation; see **Figure 6b**). The K isotopes of the different groups of CCs can be explained to the first order by the percentages of chondrule and matrix components, and to the second order, by the variation of chondrule sizes (corresponding to the different degree of K loss during chondrule formation; Hewins et al. 1997). It has been shown that chondrule sizes and K concentrations vary significantly between different groups of chondrites (Rubin, 2000; Jones, 2012; Hezel et al., 2018a). Regardless of whether the chondrule size distributions in the different CC groups were caused by physical size-sorting in turbulent nebula (*e.g.*, Cuzzi et al. 2000) or due to repeated melting in different nebular environments (Rubin, 2010), the K isotopes of chondrules were likely fractionationed during their initial melting or remelting.

We note that chondrule sizes (or the degree of the loss of K during chondrule formation) also vary within the same meteorite (*e.g.*, Rubin 1989), so some of the scattering of K isotopes in this study may also be due to the smaller sample sizes (~100 mg) where larger chondrules (up to 10 mg each), or a lack there of, may skew the K isotope data of the bulk meteorites analyzed here. This could also explain some of the observed heterogeneity seen between different fragments of the same meteorite (see **Table 1**).

There is one outlier (CV3 LAR 12002) that cannot be explained by this mixing model. We analyzed two fragments of this meteorite independently and the results are similar (−1.45 ± 0.04 ‰ vs. −1.54 ± 0.03 ‰). This sample is distinct from the other two

CV3 chondrites (Allende and GRA 06101), which are similar to each other and agree with the model. LAR 12002 is also an outlier among all 17 CCs, in that is the only CC that shows a K isotope composition that is significantly lower than the BSE value ($\sim -0.5$ ‰; see **Figure 2**). All other CCs have K isotope compositions that are similar to or higher than the BSE value (see **Figure 2**). Currently, no detailed study of this sample has been conducted, so it is possible that the K isotopic value of LAR 12002 has been altered by terrestrial weathering and/or contamination. For example, although GRA 06101 has been assigned the weathering grade A/B, abundant white greenish-blue evaporites were found on the surface of this sample (Satterwhite, 2013). Previous studies of other Antarctic meteorites have shown that these evaporites contain K (Gooding, 1981; Velbel et al., 1991). Nevertheless, it is unknown whether something such as these K-bearing evaporites affected the K isotope composition of LAR 12002, so further study is needed to understand its unique K isotope composition. As of now, we treat it as an outlier among the CCs due to its possible K isotopic fractionation during terrestrial weathering.

In summary, the K isotope fractionation observed in the CCs can be explained by a combination of both nebular and parent-body processes. The parent-body modification (aqueous and thermal alteration) tends to reduce the K isotope variation shown in Type 3 chondrites. The main pattern of K isotope variation among all CC groups, however, can generally be explained with the four-component mixing model (Alexander, 2019). While Alexander (2019)'s model assumed only one set (average) of values for the elemental compositions and isotopic anomalies of the chondrule component, here, we require a range of K isotope compositions for the chondrule component in order to explain the variation of K isotopes observed in the bulk CCs. This requirement is consistent with the the chondrule size distributions observed in the different chemical groups of chondrites (Rubin, 2000; Jones, 2012) and the chondrule size correlation with K concentration/isotopes (Hewins et al., 1997). As such, this suggests the importance of physical size-sorting of particles in the nebula on the chemical and isotopic fractionation of different groups of CCs in addition to the mixing of different chemical and isotopic reservoirs (*e.g.*, Alexander 2019).

## 4.3 Potassium isotope fractionation of ordinary chondrites

The OCs in this study include all three chemical groups, and petrological types, as well as a wide range of shock levels, cosmic-ray exposure ages, both falls and finds, and extents of terrestrial weathering. As discussed in **Section 4.1** and the **Online Supplementary Material** (see **Figure 3** and **Supplementary Figure S1**), we observed no K isotopic variation correlating with either falls or finds, degree of terrestrial weathering, or cosmic-ray exposure ages. Below, we discuss whether chemical group, petrological type or shock level resulted in any measurable effect on the K isotopic compositions of OCs.

*Chemical groups*: Unlike the CCs, there is no obvious trend of lithophile MVE depletion versus their nebular volatilities among the three chemical groups (see **Figure 1**). Compared to the CCs, the three OC chemical groups, in general, have similar abundances of MVEs (see **Figure 1**). In this study, we show that the K isotopic composition of the three chemical groups also have similar and comparable ranges (see **Figure 2**). The 28 samples studied here cover all three chemical groups, with 10 samples belonging to the H group, 10 belonging to the L group, and 8 belonging to the LL group. The three different chemical groups of H, L, and LL represent at least three different parent bodies (Krot et al., 2014). Although all three groups exhibit a similar (~1 ‰) range of K isotope compositions (from −1.5 ‰ to −0.5 ‰) and all generally have lighter K isotope compositions than the BSE (−0.5 ‰), the H group has two samples that are significantly enriched in heavy K isotopes compared to the BSE and other OCs (see **Figure 2b**). These two samples are RKPA 78004 (H4, 0.00 ± 0.03 ‰) and PRE 95400 (H5, 0.07 ± 0.03 ‰ and 0.04 ± 0.05 ‰ for the two fragments measured here). Excluding these two samples, the average K isotope composition of the H group is not significantly different from those of the L and LL groups. Thus, currently there appears to be no bulk K isotopic differences between the H, L and LL parent bodies.

*Thermal metamorphism*: Petrologic Types 4 to 6 represent samples that have experienced increasing degrees of thermal metamorphism and higher degrees of chemical equilibration in their parent bodies. **Figure 5 (b, c and d)** shows K isotopes versus the petrologic types for the H, L and LL groups. We found no apparent correlation between the K isotopes and petrologic types within each group. The Type 4 and 5 H groups have the largest variations; however, again this is simply due to the two outliers of RKPA 78004 (H4) and PRE 95400 (H5) (see **Figure 5b**). Excluding these two outliers, the range of K isotope compositions of each petrological type from each chemical group generally overlap with each other and show no consistent trends.

As discussed in **Section 4.2**, large (a few ‰) K isotopic fractionations have been observed in chondrules, so consequently, the variation of K isotopes in "bulk" chondrites to some degree depends on the chondrule modal abundances in the specific fragments of the meteorites studied (e.g., different chips of the same meteorites; see **Table 2**). Type 3 OCs would thus be expected to have the largest variation/heterogeneity in K isotopes. With increasing thermal metamorphism, K becomes more mobile and chondrules would exchange K with other chondrules and the matrix (Grossman and Brearley, 2005). The thermal metamorphism in parent bodies would be expected to equilibrate the K isotopes between different components (*i.e.*, chondrule-chondrule and chondrule-matrix), which have very different initial K isotope compositions (Alexander et al., 2000; Alexander and Grossman, 2005). Thus, with increasing petrologic type, OCs would be expected to have less variation in K isotopes than the unequilibrated Type 3s. Nevertheless, due to the limited number of samples analyzed for K isotopes for each petrological type, it is still not clear whether the expected K isotope homogenization occurred in OCs or not.

*Aqueous alteration and metasomatism*: Although the petrological type (3-6) usually assigned to OCs does not indicate the existence and degree of aqueous alteration like the petrological type (1-2), evidence of aqueous alteration and metasomatism has been widely observed in both equilibrated and unequilibrated OCs (e.g., Brearley, 2007; Lewis and Jones, 2016; Lewis and Jones, 2019). For example, Hutchison et al. (1987) and Alexander et al. (1989) reported the presence of smectite and calcite in Type 3 OCs, which provided the first direct evidence of in-situ aqueous alteration. In addition, Grossman et al. (2000) described textually distinct and alkli-depleted chondrules within OCs of all petrological types and interpreted these as water bleached chondrules. As another example, aqueous fluid inclusions in alkali chlorides (halide NaCl and sylvite KCl) were found in the matrix of the H5 OC Monahans (Zolensky et al., 1999). Potassium is soluble, and it has recently been reported that dissolved K in solution (*e.g.*, river water) is usually enriched in heavier isotopes when compared to the leached residue (Li et al., 2019). Thus, aqueous alteration in parent bodies could potentially change the pre-accretionary K isotopic compositions and fractionate K isotopes between different components of chondrites (*e.g.*, bleached chondrules vs. matrix containing precipitated salts, as well as clay minerals and hydrothermal feldspars). We would expect the bleached chondrules to be enriched in the lighter K isotope, while the K salts (precipitated from the bleaching water; Rubin et al. 2002; Zolensky et al. 1999) in the matrix would be enriched in the heavier K isotope. However, such a prediction cannot be confirmed with the *bulk* analyses of OCs and so would need to be tested with future *in-situ* studies.

The aqueous alteration in the OCs has been found to be decoupled from thermal metamorphism (Grossman et al., 2000). Therefore, petrological type (3-6) is not a good indicator for the degree of aqueous alteration in OCs. Currently there is no adequate index for such a purpose. Thus, it is challenging to evaluate aqueous alteration effects on the K isotopes of OCs even in a qualitative fashion. For example, Grossman et al. (2000) found that L and LL OCs contain more bleached chondrules than H OCs, meaning L and LL OCs have likely been more aqueously altered than H OCs. However, as shown in **Figure 5**, we do not observe any significant difference in K isotopes between H and L/LL OCs. We should note, however, that although K is mobile during aqueous alteration and the K can be leached out from chondrules and deposited in the matrix, large and representative *bulk* chondrites should be treated as a closed-system for K (not for $H_2O$ as it could be lost from the parent body due to thermal metamorphism) (Grossman et al., 2000). This might be the reason why no systematic K isotopic differences are observed between bulk H and L/LL OCs. Nevertheless, aqueous alteration, in principle, would be a possible mechanism for generating K isotope fractionation and heterogeneity in the parent bodies of the OCs. The two outliers, RKPA 78004 (H4) and PRE 95400 (H5), could possibly have been affected by aqueous alteration, which would make the fragments studied here not representative of the bulks. However, no detailed studies of aqueous alteration in these two samples are currently available.

*Impact vaporization*: As previously discussed in **Section 4.2**, impact vaporization is another mechanism that is known to fractionate the isotopes of MVEs such as Zn and Cu (*e.g.*, Moynier et al., 2009; Frederic Moynier et al., 2010). However, for OC samples from the same chemical groups and petrological types, we did not observe any correlation between shock stage and K isotope composition (see **Table 1**). For example, Bjurböle (shock level S1) and ALH 85033 (shock level S6) are both L4 OCs which experienced very different degrees of shock yet they have indistinguishable K isotopic compositions ($-0.83 \pm 0.05$ ‰ vs. $-0.92 \pm 0.08$ ‰). ALH 85017 (shock level S6) and ALHA 76001 (shock level S3) are another example, as both are L6 finds with different degrees of shock but show very similar K isotope compositions ($-0.74 \pm 0.03$ ‰ vs. $-0.82 \pm 0.04$ ‰). This observation is consistent with a recent thermochemical calculation that shows, in general, a lower volatility of K compared to other MVEs such as Zn and Cu during impact vaporization (Jiang et al., 2019). Their observation on the K isotopes of tektites (impact-generated melts formed on Earth) shows a similar trend to what is seen in OCs here, with no K isotope fractionation compared to the source rocks (in this case of tektites) (Jiang et al., 2019), in contrast to other MVE isotopes such as Zn and Cu, which show significant fractionation (Moynier et al., 2009; Moynier et al., 2010a; Moynier et al., 2010b). Thus, unlike Zn and Cu isotopes, the K isotope variation among OCs observed here cannot be explained as the result of impact vaporization.

In summary, the K isotope fractionation observed in OCs can be generated by both nebular and parent-body processes. Nevertheless, there are no significant trends or differences between the different OC chemical groups or petrologic types and there are often larger variations within the same groups than there is between groups. Consequently, we attribute the full range of K isotopic compositions observed in the OCs to the combined effects of 1) K isotope fractionation during chondrule formation as suggested by measurments of individual chondrules from OCs (Alexander et al., 2000; Alexander and Grossman, 2005); 2) mixing of different sized chondrules; 3) chondrules and matrix propotions; 4) parent-body aqueous and thermal alteration; and 5) sampling bias.

## 4.4 Comparison of K isotopes with isotopes of other moderately volatile elements (Cu, Zn, and Rb) in chondrites

Potassium is a moderately volatile element with a 50 % condensation temperature of 1006 K (Lodders, 2003), which is similar to the 50 % condensation temperatures of Cu (1037 K). Zinc (726 K) and Rb (800 K) have 50 % condensation temperatures that are lower than K; however, they all belong to the MVE group and are often compared to each other. Here, we compare our chondrite K isotope dataset with previously published chondrite isotope data for these other MVEs (Cu, Zn, and Rb).

Among the three MVEs listed above, Zn isotopes are the most extensively studied, with the Zn isotopes of CCs and OCs having been measured by Luck et al. (2005), Moynier et al. (2007), Barrat et al. (2012), and Pringle et al. (2017). These studies found a positive correlation between Zn isotopes and Zn concentrations among the CCs, where CI chondrites have the highest Zn concentration and the heaviest Zn isotopic composition. This positive correlation among the different CC chemical groups is the opposite to what was expected for Zn isotope fractionation via a simple kinetic evaporation. Thus, Luck et al. (2005) and Pringle et al. (2017) interpreted this positive trend as the mixing between two isotopically distinct reservoirs (high Zn and heavier Zn isotope reservoir vs. low Zn and lighter Zn isotope reservoir). As for OCs, all three groups show similar ranges of Zn isotope variation and, in general, are enriched in lighter Zn isotopes in comparison CCs (Luck et al., 2005). As shown in **Figure 4c**, there is no clear trend between K isotopes and K concentrations. As discussed in **Section 4.2**, for the CCs our K isotopes and K concentrations can be satisfactorily explained by Alexander (2019)'s four-component-mixing (chondrule, refractory inclusion, matrix and water) model. However, for simplicity and ease of fitting Alexander (2019)'s model assumes a complete loss of Zn from the chondrule component and no Zn in the refractory inclusions or water, so all the Zn would have to come from the single source/reservoir: the matrix. This would cause all CCs to have the same Zn isotope composition regardless of their chemical groups and Zn concentrations, which Alexander (2019) acknowledged is not consistent with the observations (Luck et al., 2005; Moynier et al., 2007; Pringle et al., 2017). As such, the Zn isotopes of chondrites cannot be reconciled with Alexander (2019)'s model and Zn isotopes also appear to be decoupled from the K isotope fractionation of chondrites seen in this study (see **Figure 7a**).

The Cu isotopes of the CCs and OCs have been analyzed by Luck et al. (2003), Moynier et al. (2007) and Barrat et al. (2012). Generally, the Cu isotope compositions of the CCs co-vary with Zn isotopes and have a positive correlation with Cu concentrations, where CI chondrites have the highest Cu concentration and the heaviest Cu isotope composition. In addition, Luck et al. (2003) identified a correlation between Cu isotopes and mass-independent O isotopes ($\Delta^{17}O$), which suggests that the Cu isotope compositions of the CCs were also created by the mixing of two distinct Cu isotope reservoirs in the solar nebula (high Cu and heavier Cu isotope reservoir vs. low Cu and lighter Cu isotope reservoir). For the OCs, the Cu isotopes fall on a different mixing line and hint that there is at least one more reservoir that is distinct from the two which make up the CCs. According to Alexander (2019)'s model, the Cu concentrations of the CCs can be derived from mixing the chondrule component with the matrix component since there is no Cu in the refractory inclusion and water components. In order to match the observed Cu isotopic compositions of the CCs (Luck et al., 2003; Moynier et al., 2007; Barrat et al., 2012), the matrix component needs to be enriched in the heaviest Cu isotope composition while the chondrule component in the lightest isotope composition. This is the opposite to what we expect for K isotopes in the chondrule and matrix components, where chondrules are enriched in heavier K isotopes due to the Rayleigh fractionation via vaporization during the chondrule formation (Yu et al., 2003; Richter et al., 2011). However, there is a possible way to reconcile the differences between the Cu and K isotopes. Copper has only two stable isotopes, thus, we cannot distinguish the Cu isotope fractionation caused by mass-dependent processes (*e.g.*, evaporation/condensation) from the variations caused by radioactive decay. If the decay of $^{63}$Ni (instead of mass-dependent processes) is dominating the Cu isotope composition, the Cu isotopes of the CCs and OCs (see details in Luck et al., 2003; Moynier et al., 2007; Barrat et al., 2012) can be consistent with the model of Alexander (2019) and be decoupled from the K isotopes at the same time (see **Figure 7b**). However, whether Cu isotopic variations among chondrites are due to the decay of $^{63}$Ni still requires further investigation.

The Rb isotopes of chondrites are the least studied and have been only reported by Pringle and Moynier (2017). They found a similar trend to the one observed for the Zn and Cu isotopes of the CCs. Again, there is a positive correlation between Rb isotopes and Rb concentrations among the CCs, whereas the CI chondrites have the highest Rb concentration and the heaviest Rb isotope composition (Pringle and Moynier, 2017). Similar to Zn and Cu isotopes, Pringle and Moynier (2017) interpreted this positive trend as the mixing between two isotopically distinct reservoirs (high Rb and heavier Rb isotope reservoir vs. low Rb and lighter Rb isotope reservoir). Only one OC has been analyzed, and its Rb isotope composition is lighter than those of all CCs. Despite the close chemical similarity between Rb and K, the Rb isotopes of chondrites cannot be reconciled with the K isotopes (see **Figure 7c**). Therefore, future experimental studies are urgently required to better our understanding of the different geochemical and cosmochemical behaviors of MVEs during nebular and/or parent-body processes.

**4.5 The K isotope comparison between carbonaceous, ordinary and enstatite chondrites**

The major and most intriguing observation of this study on 16 CCs and 28 OCs is that although both CCs and OCs overlap with the BSE value, most CCs have heavier K isotopes relative to the BSE while that majority of OCs have lighter K isotopes. This observation suggests a dichotomy in K isotopes between the CCs and OCs. A dichotomy between the CCs and OCs can be also found in the mass-dependent isotope systems of other MVEs, such as Cu, Zn and Rb (Luck et al., 2003; Luck et al., 2005; Moynier et al., 2007; Barrat et al., 2012; Wang and Jacobsen, 2016b; Pringle et al., 2017; Pringle and Moynier, 2017). In each case, although with some overlap, the CCs are typically enriched in the heavier isotopes relative to the OCs. In contrast, Li is the least volatile element among MVEs (*e.g.*, Sossi and Fegley, 2018) and there is no measurable difference between CCs and OCs in Li isotopes (Pogge von Strandmann et al., 2011). Similarly, for the mass-dependent isotope systems of non-volatile elements (50 % condensation temperatures higher than Li), CCs and OCs also show no resolvable difference, such as Mg (*e.g.*, Teng et al., 2010; Schiller et al., 2010; Bourdon et al., 2010; Pogge von Strandmann et al., 2011; Hin et al., 2017), Si (*e.g.*, Georg et al., 2007; Fitoussi et al., 2009; Armytage et al., 2011; Dauphas et al., 2015), Ca (*e.g.*, Simon and DePaolo, 2010; Valdes et al., 2014; Huang and Jacobsen, 2017), Ti (*e.g.*, Greber et al., 2017; Deng et al., 2018), V (*e.g.*, Nielsen et al., 2014; Xue et al., 2018), and Fe (*e.g.*, Craddock and Dauphas 2011; Poitrasson et al. 2004; Wang et al. 2013). The fact that only isotopes of elements more volatile than Li (*e.g.*, K, Cu, Zn and Rb) show a mass dependent dichotomy between the CCs and OCs is still intriguing. It is possible that the reservoirs or components of the OCs and CCs experienced different extents of evaporation/condensation processes or dust-gas separation that enriched heavier isotopes of MVEs in CCs relative to OCs. The K isotopic dichotomy, at least, cannot be simply explained by post-accretionary processes (e.g., aqueous and thermal alterations on parent-bodies) as discussed in **Sections 4.2** and **4.3**.

We want to clarify that this dichotomy between OCs and CCs in K isotopes does not extend to all non-carbonaceous chondrites (NCs; OCs + enstatite chondrites + Rumuruti chondrites) versus CCs. If enstatite chondrites (ECs) are included there is no dichotomy in K isotopes, or Cu, Zn and Rb isotopes, between the NCs and CCs (Moynier et al., 2007; Moynier et al., 2011; Pringle and Moynier, 2017; Zhao et al., 2019). As shown in **Figure 8**, although there is a clear separation between the probability distribution functions of K isotopes in the CCs and OCs, the ECs fall in the middle and overlap with both the CCs and OCs and cannot be distinguished from either. Although the data for the ECs are highly scattered, interestingly, among the CCs, OCs and ECs, ECs have the most similar K, Cu, Zn and Rb isotopes to the Earth (Moynier et al., 2007; Moynier et al., 2011; Pringle and Moynier, 2017; Zhao et al., 2019), which is to some degree surprising since the Earth is depleted in MVEs compared to ECs (see **Figure 1**). It is not known why the MVE isotopes were not fractionated during the process(es) that depleted the MVEs in Earth. Nevertheless, such (mass-dependent) isotopic similarity between the ECs and Earth in MVEs is consistent with the model suggesting that the ECs and Earth formed from the same precursor materials based on isotopic anomalies (e.g., Javoy, 1995; Javoy et al., 2010; Dauphas, 2017).

Lastly, we want to note that the lack of a dichotomy between the NCs and CCs in MVEs isotopes is in contrast to the clear dichotomy between the NCs and CCs that is widely observed in elemental abundances (*e.g.*, Mg/Si vs. Al/Si) and in many *mass-independent* isotope systems (Warren, 2011a; Warren, 2011b). Different from *mass-independent* isotope systems, the mass-dependent isotope systems are not the most suitable to trace the sources/precursors since they can be fractionated by many physical and chemical processes, such as evaporation/condensation in the solar nebular environments or aqueous alteration/thermal metamorphism on the parent-body asteroids.

# CONCLUSIONS

In this study, we report the most complete and systematic K isotope study of 16 CCs and 28 OCs, covering the entire spectrum of chemical groups and petrological types. We observed up to ~2 ‰ variations in the K isotopic composition of chondrites, which is consistent with the findings of Humayun and Clayton (1995). With the improved precision, we provide the best current estimate of K isotopic compositions for the CCs and OCs. The variation observed among the CCs and OCs can be explained by the combined effects of both solar nebular and parent-body processes.

We do not observe any significant correlations between the K isotopic compositions and K elemental concentrations (or elemental ratios such as K/U, K/Si and K/Al). This rules out a single-stage partial vaporization process to account for both the K isotope variations found here, as well as the well-known progressive elemental depletion pattern of moderately volatile elements among CCs. Nevertheless, both the K isotope pattern demonstrated here and the MVE depletion pattern of CCs can be explained by the mixing model of Alexander (2019), with isotope fractionation during chondrule formation and chondrule size distribution among CCs playing an important role.

Although there are no clear correlations between K isotopes and petrological types for either the CCs or OCs, based on previous abundant petrological and geochemical evidence on K mobility during aqueous alteration and thermal metamorphism, we propose that parent-body alterations potentially fractionate the K isotope compositions of both the CCs and OCs. The simple scheme of petrological types (1-6) assigned to meteorites cannot precisely describe the K mobility and isotope fractionation. Parent-body alterations would not only have contributed to the variation between meteorites, but also to the heterogeneity within the same meteorites observed in this study. However, the *bulk* asteroids should remain a closed-system for K unlike some volatiles ($H_2O$) which can be lost to space during a high degree of thermal metamorphism.

The most prominent observation of this study is the K isotopic difference between the CCs and OCs, which is consistent with observations seen by other mass-dependent isotope systems of MVEs. However, when ECs are combined with OCs as a combined NC, there is no dichotomy observed between the them and CCs in any MVE isotope system, which is in contrast to the clear dichotomy observed in mass-independent isotope systems.

# ACKNOWLEDGEMENTS

We thank the Associate Editor Dr. Thorsten Kleine for prompt handling and editing of this manuscript. We also thank reviewers Dr. Conel Alexander and Dr. Haolan Tang as well as one anonymous reviewer for their thorough comments. We would like to thank Dr. Kevin Righter from NASA Johnson Space Center, and Dr. Philipp Heck from The Field Museum in Chicago for providing samples. We would also like to thank the McDonnell Center for the Space Sciences and WUSTL Office of Undergraduate Research for funding our research endeavors. Work by KL is supported in part by NSF AST 1517541. MKP has received funding from the European Union's Horizon 2020 research and innovation programme under the Marie Sklodowska-Curie grant agreement No 753276. YJ is supported by NSFC (Grant Nos. 41873076).

US Antarctic meteorite samples are recovered by the Antarctic Search for Meteorites (ANSMET) program, which has been funded by NSF and NASA, and characterized and curated by the Department of Mineral Sciences of the Smithsonian Institution and Astromaterials Acquisition and Curation Office at NASA Johnson Space Center.

**Table 1.** The K concentration and isotopic compositions of carbonaceous chondrites in this study and from literature.

| Sample | Type | Fall /Find | Mass [mg] | K [ppm] this study | K [ppm] literature | $\delta^{41}K_{NIST}$ [‰] | 95 % c.i. [a] | n [b] | Shock Level [c] | Weathering Grade [c] | Source [d] | Literature |
|---|---|---|---|---|---|---|---|---|---|---|---|---|
| *CI* | | | | | | | | | | | | |
| Orgueil | CI1 | Fall | | | | −0.53 | 0.11 | 11 | | | | (Wang and Jacobsen, 2016b) |
| *CM* | | | | | | | | | | | | |
| ALH 83100 | CM1/2 | Find | 107.1 | 234 | 415 | −0.18 | 0.04 | 10 | | Be | ANSMET | (Jarosewich, 1990) |
| ALH 85013 #1 | CM2 | Find | 157.0 | | | −0.39 | 0.03 | 10 | | A | ANSMET | |
| ALH 85013 #2 | CM2 | Find | 101.1 | 159 | | −0.38 | 0.03 | 12 | | A | ANSMET | |
| EET 96029 | CM2 | Find | 103.6 | 192 | | −0.09 | 0.03 | 11 | | A/B | ANMSET | |
| LON 94101 | CM2 | Find | 101.4 | 290 | | −0.22 | 0.03 | 10 | | Be | ANSMET | |
| | | | | | CM average = | −0.25 | 0.26 [f] | | | | | |
| *CO* | | | | | | | | | | | | |
| MIL 090010 | CO3 | Find | 113.1 | 218 | | −0.17 | 0.02 | 11 | | A/B | ANSMET | |
| MIL 11213 #1 | CO3 | Find | 123.0 | | | 0.36 | 0.05 | 13 | | A | ANSMET | |
| MIL 11213 #2 | CO3 | Find | 108.5 | 82 | | 0.02 | 0.03 | 12 | | A | ANSMET | |
| ALH 83108 | CO3.5 | Find | | | | −0.35 | 0.07 | 19 | S1 | A | ANSMET | |
| ALHA77003 | CO3.6 | Find | 109.0 | 253 | 500 | −0.30 | 0.03 | 10 | S1 | Ae | ANSMET | (Jarosewich, 1984) |
| | | | | | CO average = | −0.09 | 0.58 [f] | | | | | |
| *CV* | | | | | | | | | | | | |
| Allende #1 | CV3 | Fall | | | | −0.34 | 0.02 | 32 | S1 | | | |
| Allende #2 | CV3 | Fall | | | | −0.53 | 0.03 | 31 | S1 | | | |
| LAR 12002 #1 | CV3 | Find | 114.5 | 539 | | −1.45 | 0.04 | 10 | | A/B | ANSMET | |
| LAR 12002 #2 | CV3 | Find | | | | −1.54 | 0.03 | 11 | | A/B | ANSMET | |
| GRA 06101 | CV3 | Find | 106.4 | 167 | | −0.30 | 0.02 | 10 | | B | ANSMET | |
| | | | | | CV average [g] = | −0.39 | 0.25 [f] | | | | | |
| *CR* | | | | | | | | | | | | |
| GRA 06100 #1 | CR2 | Find | 111.0 | | | 0.16 | 0.03 | 16 | | B | ANSMET | |
| GRA 06100 #2 | CR2 | Find | 113.8 | 216 | | 0.70 | 0.02 | 9 | | B | ANSMET | |
| | | | | | CR average = | 0.43 | 0.77 [f] | | | | | |
| *CK* | | | | | | | | | | | | |
| ALH 85002 #1 | CK4 | Find | 108.0 | | 247 | −0.07 | 0.02 | 16 | S2 | A | ANSMET | (Kallemeyn et al., 1991) |
| ALH 85002 #2 | CK4 | Find | 109.7 | 407 | 247 | −0.06 | 0.04 | 12 | S2 | A | ANSMET | (Kallemeyn et al., 1991) |
| ALH 85002 #3 | CK4 | Find | | | 247 | −0.11 | 0.05 | 9 | S2 | A | ANSMET | (Kallemeyn et al., 1991) |
| EET 92002 | CK5 | Find | 109.5 | 185 | | −0.15 | 0.04 | 9 | | A/Be | ANSMET | |
| RBT 03522 | CK5 | Find | 122.0 | 198 | | −0.36 | 0.03 | 11 | | B | ANSMET | |
| | | | | | CK average = | −0.15 | 0.25 [f] | | | | | |
| *CH* | | | | | | | | | | | | |
| PCA 91467 | CH3 | Find | 104.3 | 132 | | −0.64 | 0.03 | 10 | S1 | B/C | ANSMET | |
| | | | | | CC average [h] = | −0.29 | 0.34 [f] | | | | | |

*Geostandard*

| BHVO−2 | −0.47 | 0.03 | 9 | USGS |
| --- | --- | --- | --- | --- |

**Table 2.** **The K concentration and isotopic compositions of ordinary chondrites in this study and from literature.**

| Sample | Type | Fall /Find | Breccia tion[a] | Mass [mg] | K [ppm] this study | K [ppm] literature | $\delta^{41}K_{NIST}$ [‰] | 95 % c.i. [b] | n[c] | Shock Level[d] | Weathering Grade[d] | Source[e] | Literature |
|---|---|---|---|---|---|---|---|---|---|---|---|---|---|
| *H* | | | | | | | | | | | | | |
| MET 00526 | H3.05 | Find | | | | | −0.88 | 0.05 | 8 | S2 | B/C | ANMSET | |
| ALHA 77299 | H3.7 | Find | | | | 640-1330 | −0.34 | 0.05 | 14 | | A | ANSMET | (Fulton and Rhodes, 1984; Brearley et al., 1989) |
| Ochansk #1 | H4 | Fall | | | | 660-858 | −1.08 | 0.05 | 13 | | | The Field Museum (ME 1443#15) | (Easton and Elliott, 1977; Leya et al., 2001) |
| Ochansk #2 | H4 | Fall | | 121.5 | 784 | 660-858 | −1.00 | 0.04 | 10 | | | The Field Museum (ME 1443#15) | (Easton and Elliott, 1977; Leya et al., 2001) |
| Ochansk #3 | H4 | Fall | | | | 660-858 | −1.01 | 0.05 | 11 | | | The Field Museum (ME 1443#15) | (Easton and Elliott, 1977; Leya et al., 2001) |
| RKPA 78004 | H4 | Find | | | | | 0.00 | 0.03 | 14 | S4 | A | ANSMET | |
| Weston #1 | H4 | Fall | Yes | 107.6 | 316 | 830 | −0.82 | 0.04 | 12 | | | The Field Museum (ME 1835#6) | (Fulton and Rhodes, 1984) |
| Weston #2 | H4 | Fall | Yes | | | 830 | −0.54 | 0.04 | 13 | | | The Field Museum (ME 1835#6) | |
| ALHA77279 | H5 | Find | | | | | −0.80 | 0.06 | 11 | | A | ANSMET | |
| ALH 84069 #1 | H5 | Find | | | | | −0.76 | 0.04 | 8 | | A | ANSMET | |
| ALH 84069 #2 | H5 | Find | | 108.0 | 480 | | −0.80 | 0.05 | 10 | | A | ANSMET | |
| Jilin #1 | H5 | Fall | No | 108.5 | 632 | 660-792 | −1.18 | 0.02 | 14 | S3 | | The Field | (Kallemeyn et |

| | | | | | | | | | | | | Museum (ME 3.53#2) | al., 1989; Yanai et al., 1995 |
|---|---|---|---|---|---|---|---|---|---|---|---|---|---|
| Jilin #2 | H5 | Fall | No | | | 660-793 | −1.28 | 0.05 | 12 | S3 | | The Field Museum (ME 3.53#2) | (Kallemeyn et al., 1989; Yanai et al., 1995) |
| PRE 95400 #1 | H5 | Find | | | | | 0.07 | 0.03 | 10 | | A | ANSMET | |
| PRE 95400 #2 | H5 | Find | | 106.9 | 613 | | 0.04 | 0.05 | 10 | | A | ANSMET | |
| Guareña #1 | H6 | Fall | | | | 750-974 | −0.68 | 0.09 | 9 | S1 | | | (Jarosewich and Brian, 1969; Nakamura, 1974) |
| Guareña #2 | H6 | Fall | | | | 750-975 | −0.67 | 0.07 | 8 | S1 | | | (Jarosewich and Brian, 1969; Nakamura, 1974) |
| QUE 90203 | H6 | Find | | | | | −0.95 | 0.06 | 11 | | A | ANSMET | |
| | | | | | | H average = | −0.70 | 0.81 [f] | | | | | |
| **L** | | | | | | | | | | | | | |
| QUE 97008 | L3.05 | Find | | | | | −0.85 | 0.02 | 8 | S2 | A | ANSMET | |
| GRO 06054 #1 | L3.6 | Find | | | | | −0.87 | 0.03 | 8 | | | ANSMET | |
| GRO 06054 #2 | L3.6 | Find | | 107.3 | 492 | | −1.01 | 0.04 | 7 | | | ANSMET | |
| GRO 06054 #3 | L3.6 | Find | | | | | −0.89 | 0.05 | 10 | | | ANSMET | |
| ALH 85033 | L4 | Find | | | | | −0.92 | 0.08 | 9 | S6 | A/B | ANSMET | |
| Bjurböle | L/LL4 | Fall | Yes | 104.6 | 530 | 820-1000 | −0.83 | 0.05 | 12 | S1 | | The Field Museum (ME 1428#14) | (Von Michaelis et al., 1968; Easton and Elliott, 1977) |
| GRO 95530 | L5 | Find | | | | | −0.72 | 0.06 | 12 | | A | ANSMET | |
| Homestead | L5 | Fall | Yes | 102.2 | 796 | 1080 | −1.14 | 0.03 | 12 | S4 | | The Field | (Jarosewich and Dodd, |
| #1 | | | | | | | | | | | | Museum (ME 313#13) | 1985) |
| Homestead #2 | L5 | Fall | Yes | | | 1080 | −0.87 | 0.04 | 12 | S4 | | The Field Museum (ME 313#13) | (Jarosewich and Dodd, 1985) |
| Knyahinya | L/LL5 | Fall | Yes | 121.2 | 661 | 830-934 | −0.93 | 0.05 | 12 | S3 | | The Field Museum (ME 1823#8) | (Kallemeyn et al., 1989; Jarosewich, 1990) |
| LAR 06303 | L5 | Find | | 133.0 | | | −0.78 | 0.04 | 9 | | A | ANSMET | |
| ALH 85017 | L6 | Find | | 119.0 | | | −0.74 | 0.03 | 17 | S6 | A | ANSMET | |
| ALHA 76001 | L6 | Find | No | 120.0 | | 910-913 | −0.82 | 0.04 | 9 | S3 | A | ANSMET | (Haramura et al., 1983; Yanai et al., 1995) |
| Bruderheim #1 | L6 | Fall | No | | | 819-1080 | −0.57 | 0.11 | 8 | S4 | | | (Baadsgaard et al., 1961; Nakamura, 1974; Wang and Jacobsen, 2016b) |
| Bruderheim #2 | L6 | Fall | No | | | 819-1080 | −0.60 | 0.10 | 8 | S4 | | | (Wang and Jacobsen, 2016b) |
| Peace River | L6 | Fall | | | | | −0.54 | 0.10 | 14 | | | | (Wang and Jacobsen, 2016b) |
| | | | | | | L average = | −0.82 | 0.32 [f] | | | | | |
| **LL** | | | | | | | | | | | | | |
| ALHA 76004 | LL3.2/3.4 | Find | | 99.0 | | 830-910 | −1.13 | 0.05 | 9 | | A | ANSMET | (Jarosewich, 1990; Yanai et al., 1995) |
| ALHA 77278 | LL3.7 | Find | | 177.0 | | 830-911 | −0.97 | 0.02 | 9 | | A | ANSMET | |
| GRO 95552 #1 | LL4 | Find | | 100.0 | | | −1.41 | 0.03 | 10 | S3 | A | ANSMET | |
| GRO 95552 #2 | LL4 | Find | | | | | −1.47 | 0.06 | 12 | S3 | A | ANSMET | |
| Hamlet | LL4 | Fall | No | | | 600 | −1.04 | 0.08 | 11 | S3 | | | (Wiik, 1969) |
| LAP 02207 | LL5 | Find | | | | | −0.44 | 0.05 | 10 | | A | ANSMET | |

| | | | | | | | | | |
|---|---|---|---|---|---|---|---|---|---|
| RBT 04127 #1 | LL5 | Find | 177.0 | | −0.84 | 0.02 | 8 | | ANSMET |
| RBT 04127 #2 | LL5 | Find | 101.3 | 532 | −0.67 | 0.04 | 10 | | ANSMET |
| PRA 04405 | LL6 | Find | | | −0.72 | 0.05 | 11 | A | ANSMET |
| PRA 04422 #1 | LL6 | Find | 98.0 | | −0.86 | 0.03 | 16 | A | ANSMET |
| PRA 04422 #2 | LL6 | Find | 111.5 | 524 | −0.94 | 0.03 | 7 | A | ANSMET |
| | | | | LL average = | −0.95 | 0.61 [f] | | | |
| | | | | OC average = | −0.81 | 0.64 [f] | | | |

**Geostandard**

| | | | | | | | | | |
|---|---|---|---|---|---|---|---|---|---|
| BHVO-2 | | | | | −0.49 | 0.03 | 11 | | USGS |

**Figure Captions:**

30    **Figure 1**. The ratios of K/Si, Cu/Si, Rb/Si and Zn/Si in different chemical groups of chondrites and bulk planetary bodies (Earth, Mars, Moon, and Vesta) normalized to those ratios in CI chondrites. The ratios of the solar photosphere are also shown. All data are from the compilation by Lodders and Fegley (1998; 2011).  The volatilities of the elements are expressed as their 50 % nebular condensation temperatures (Lodders, 2003).

**Figure 2**. The bulk K isotopic compositions of carbonaceous chondrites (A) and ordinary chondrites (B) in this study and literature (Wang and Jacobsen, 2016b).

**Figure 3**. The bulk K isotopic compositions of carbonaceous chondrites (A) and ordinary
40    chondrites (B) vs. the weathering grades. The dashed line represents the BSE value (−0.5 ‰).

**Figure 4**. K isotopic compositions versus K/U (ppm/ppm; normalized to CI) and K/Al (ppm/ppm; normalized to CI) ratios and K abundance (ppm; normalized to CI) of carbonaceous and ordinary chondrites samples from this study and literature (Jarosewich,
45    1984; Fulton and Rhodes, 1984; Jarosewich and Dodd, 1985; Kallemeyn et al., 1989; Kallemeyn et al., 1991; Yanai et al., 1995; Wang and Jacobsen, 2016b). The analytical errors are smaller than the sizes of the symbols if not shown. The dash line represents the BSE value (−0.5 ‰).

**Figure 5**. Range of K isotopic compositions of carbonaceous chondrites (A), H
50    chondrites (B), L chondrites (C) and LL chondrites (D) vs. petrological types. The middle line represents the median of the data. The box extends from the 25[th] to 75[th] percentiles of the data. The bars mark the maximum and minimum values of each petrological type. The triangles indicate outlier. The dashed line represents the BSE value (−0.5 ‰). N
55    represents the number of samples (including the different fragments/chips of the same meteorites in order to show the sample heterogeneity).

**Figure 6**. Mixing model of K concentrations (A) and K isotopic compositions (B) of carbonaceous chondrites based on Alexander (2019). The K concentrations normalized to
60    Si and CI are from compilation by Alexander (2019). The K isotope data are from this study and literature (Wang and Jacobsen, 2016b). The percentages indicate the percentages of K loss during chondrule formation based on our Rayleigh modeling (see **Discussion Section 4.2**).

**Figure 7**. K isotope compositions versus Zn isotope (A), Cu isotope (B), and Rb isotope (C) compositions of carbonaceous and ordinary chondrites from this study and literatures (Luck et al., 2003; Luck et al., 2005; Moynier et al., 2007; Barrat et al., 2012; Wang and Jacobsen, 2016b; Pringle et al., 2017; Pringle and Moynier, 2017). The error bars for Zn, Cu, and Rb isotopes are the reported analytical uncertainties (2SD) from the literatures.

**Figure 8**. Histograms and relative Probability Density Function (PDF) of K isotopic compositions of carbonaceous chondrites, ordinary chondrites, and enstatite chondrites from this study and literature (Wang and Jacobsen, 2016b; Zhao et al., 2019). The dash line represents the BSE value (−0.5 ‰). N represents the number of meteorites (not including the different fragments/chips of the same meteorites).

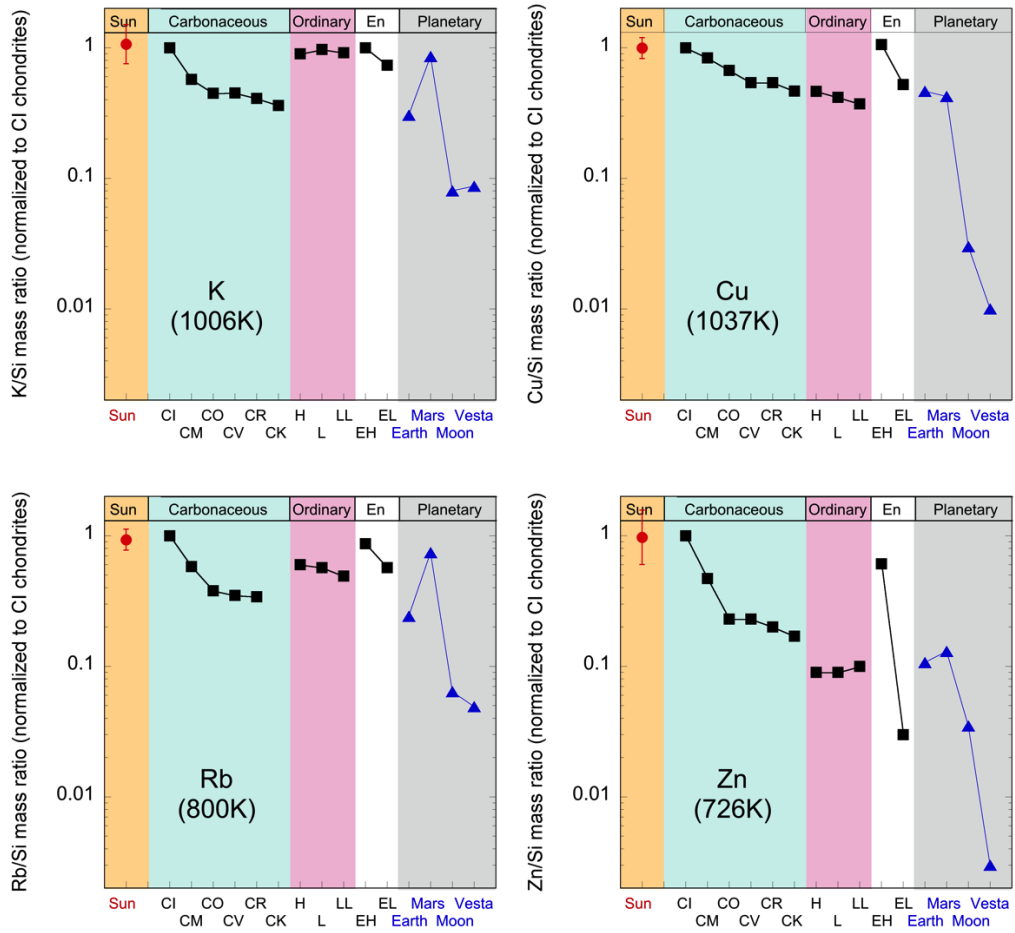

**Figure 1**

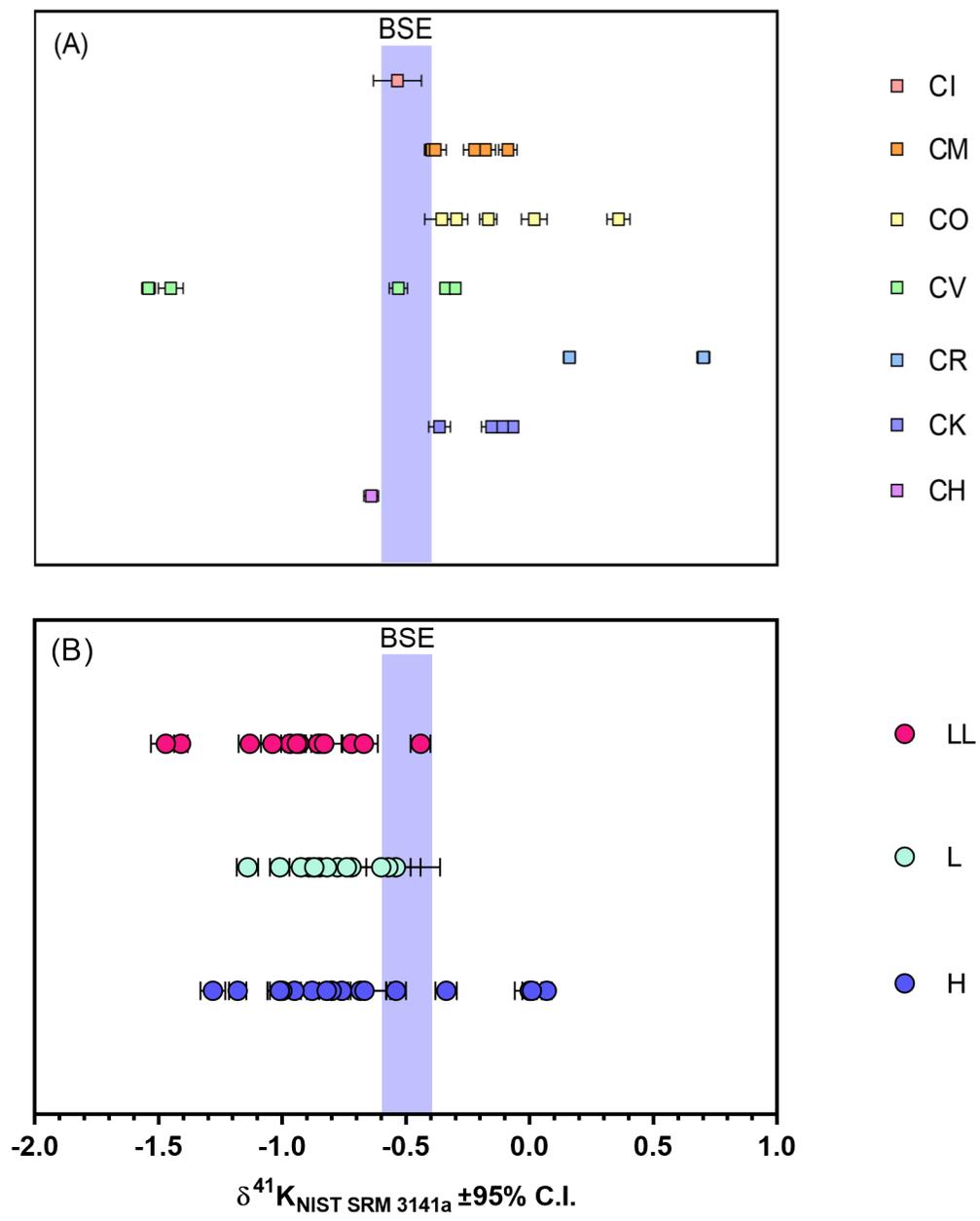

**Figure 2**

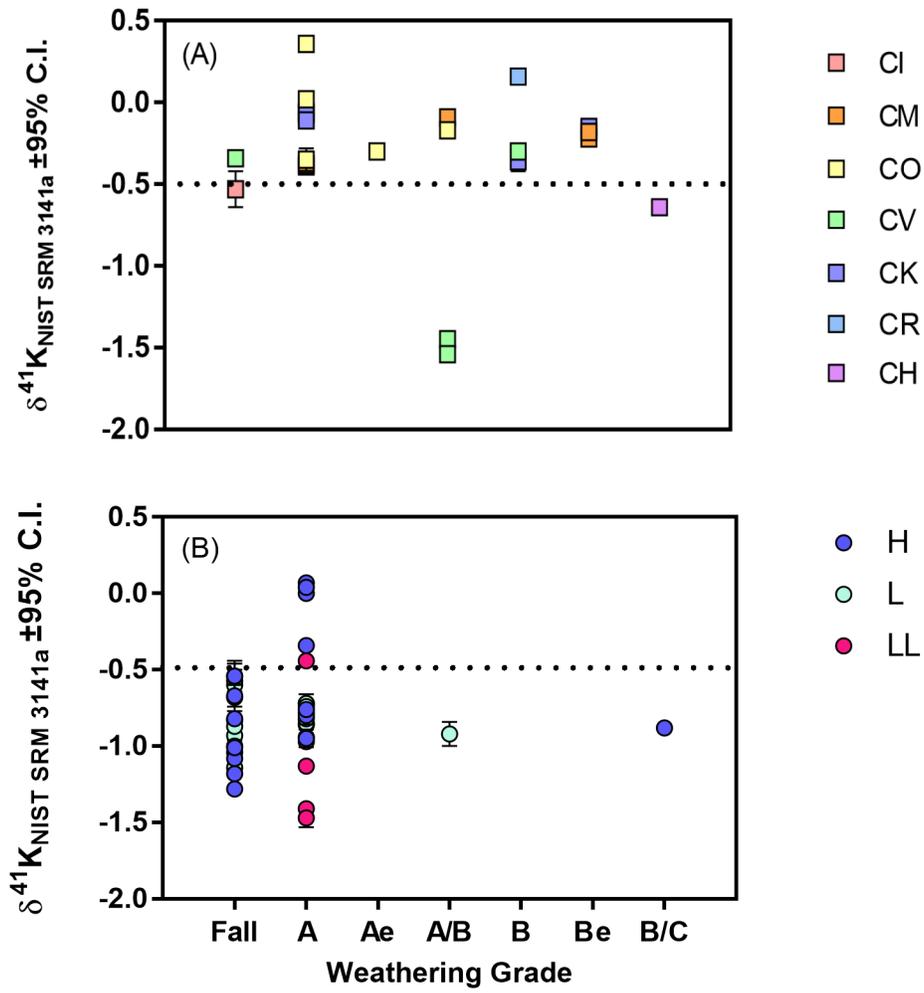

**Figure 3**

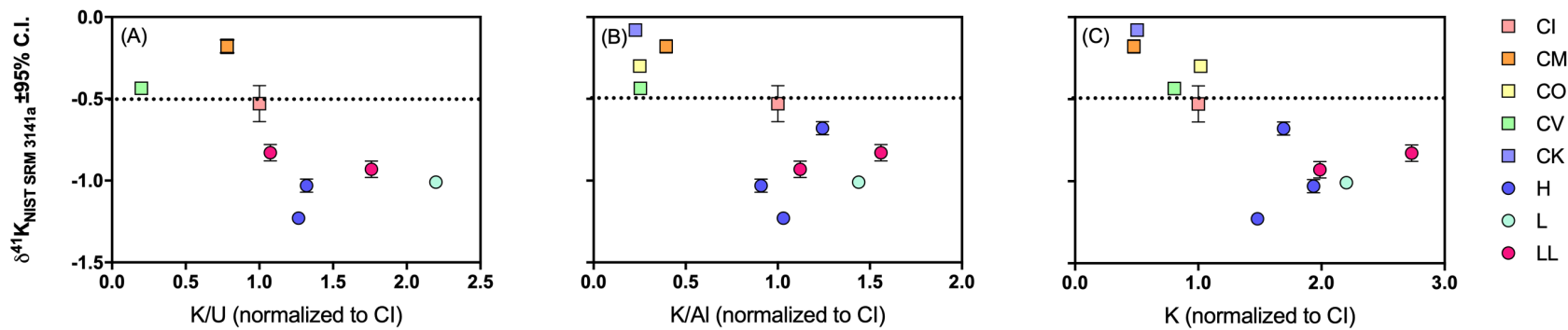

**Figure 4**

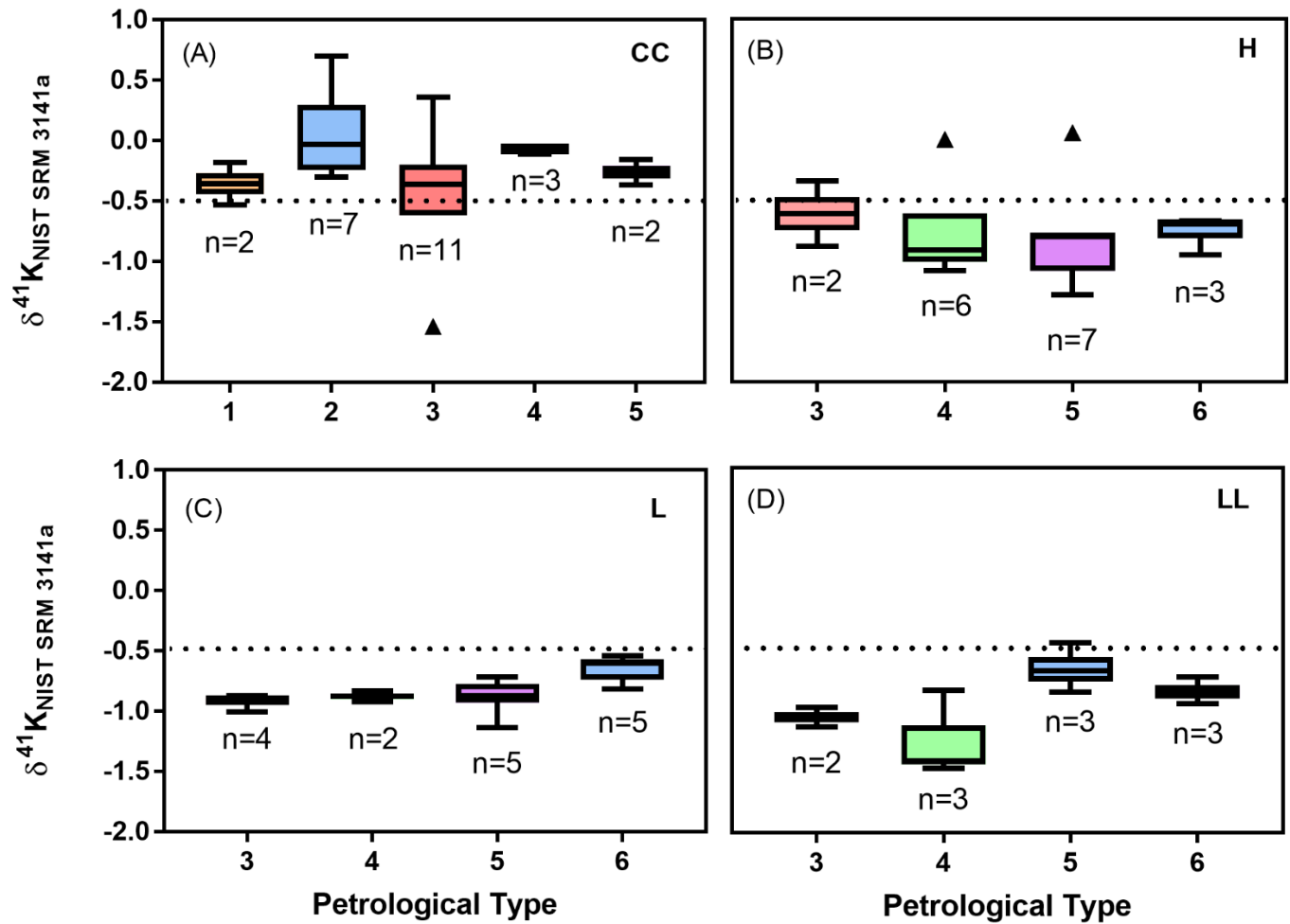

Figure 5

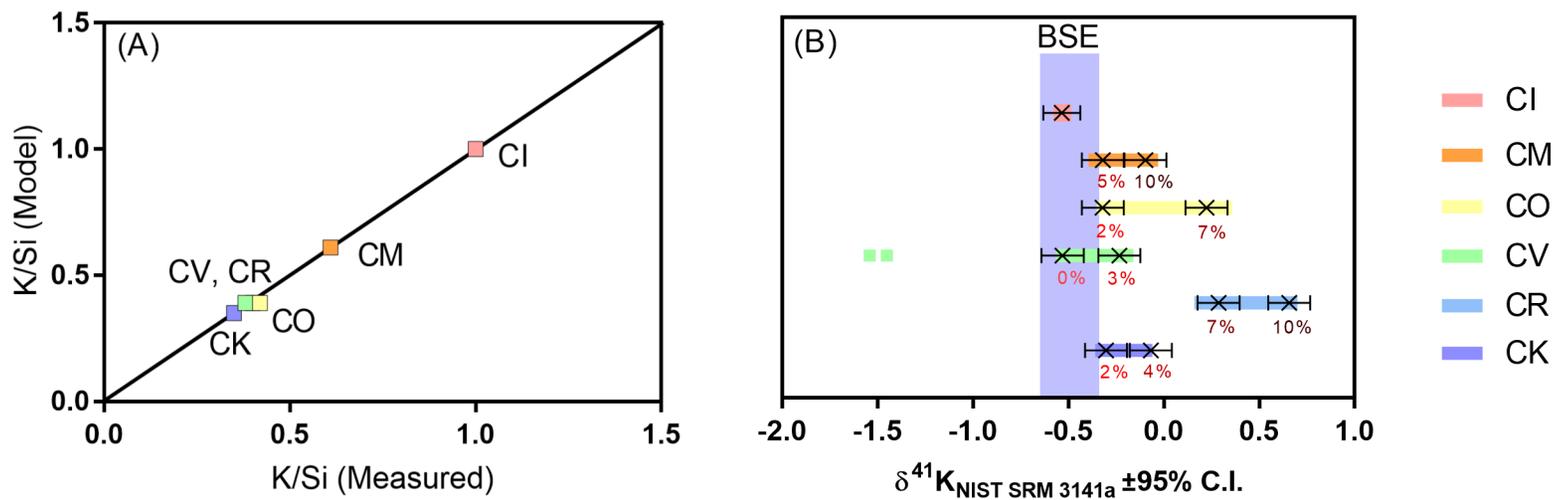

**Figure 6**

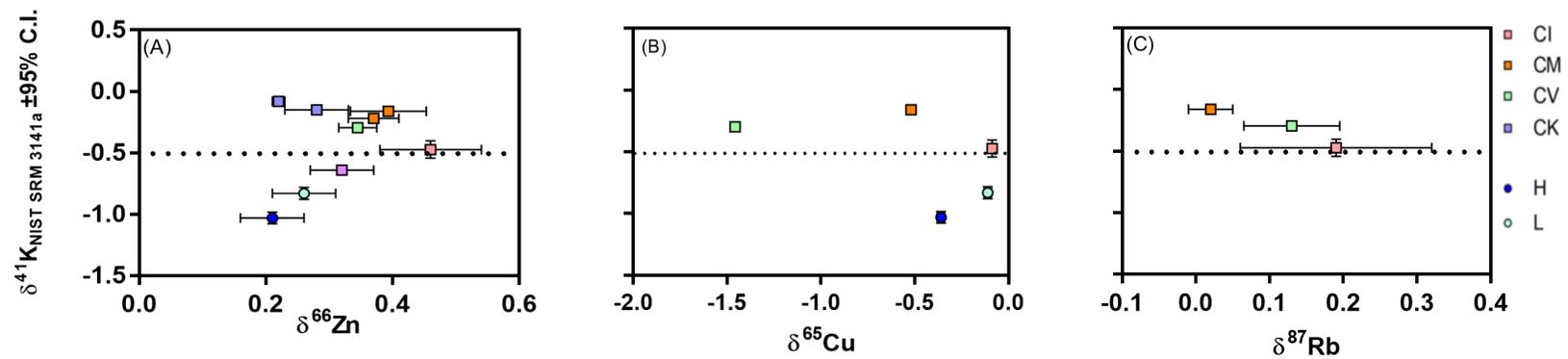

**Figure 7**

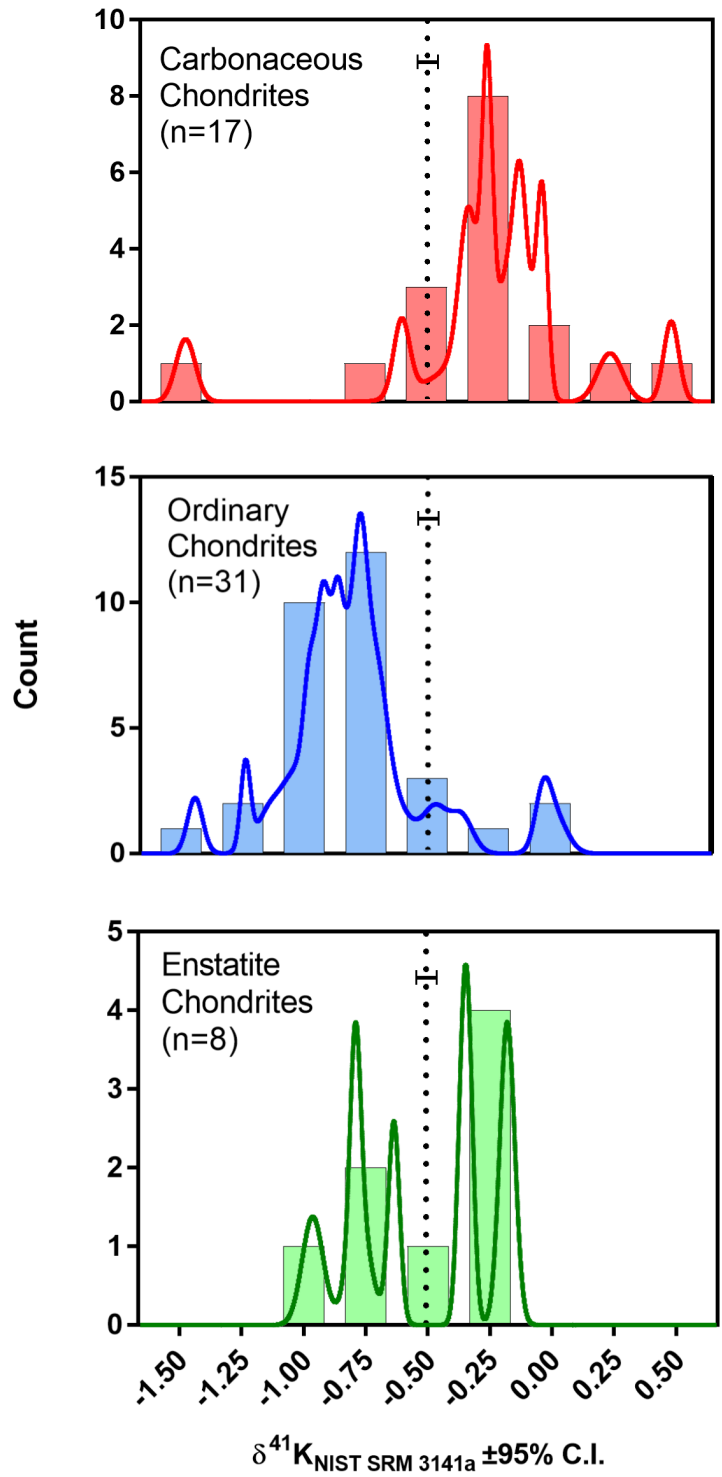

**Figure 8**

**Online Supplementary Material**

**for**

**Potassium Isotope Compositions of Carbonaceous and Ordinary Chondrites: Implications on the Origin of Volatile Depletion in the Early Solar System**

Hannah Bloom[1], Katharina Lodders[1], Heng Chen[1,2], Chen Zhao[1,3], Zhen Tian[1], Piers Koefoed[1], Mária K. Pető[4], Yun Jiang[5,6], and Kun Wang (王昆)[1]*

[1]Department of Earth and Planetary Sciences and McDonnell Center for the Space Sciences, Washington University in St. Louis, One Brookings Drive, St. Louis, MO 63130, USA

[2]Lamont-Doherty Earth Observatory, Columbia University, Palisades, NY 10964, USA

[3]Faculty of Earth Sciences, China University of Geosciences, Wuhan, Hubei 430074, China

[4]Konkoly Observatory, Research Center for Astronomy and Earth Sciences, H-1121 Budapest, Hungary

[5]CAS Key Laboratory of Planetary Sciences, Purple Mountain Observatory, Chinese Academy of Sciences, Nanjing 210008, China

[6]CAS Center for Excellence in Comparative Planetology, China

*Corresponding author email: wangkun@wustl.edu

Accepted version

**Geochimica et Cosmochimica Acta**

**How to cite:**

Bloom, H., Lodders, K., Chen, H., Zhao, C., Tian, Z., Koefoed, P., Pető, M. K., Jiang, Y., and Wang, K. (2020) Potassium isotope compositions of carbonaceous and ordinary chondrites: Implications on the origin of volatile depletion in the early solar system. *Geochimica et Cosmochimica Acta*, **in press.**

**Effect of cosmic-ray radiation on potassium isotopes of chondrites.**

It is well known that meteorites have experienced galactic and solar cosmic-ray radiation while traveling through space. The cosmic-ray-induced spallation reactions can change the isotope compositions of K in meteorites, particularly in iron meteorites (*e.g.*, Voshage 1978). All three isotopes of K ($^{39}$K, $^{40}$K, and $^{41}$K) can be generated via interaction of cosmic rays and target elements (mainly Fe and Ni) and the $\delta^{41}$K range of iron meteorites could vary by as much as 1,000 ‰ (Voshage et al., 1983). Therefore, the $^{41}$K/$^{40}$K-$^3$He/$^{21}$Ne radiation exposure dating method has been applied routinely to iron meteorites (*e.g.*, Voshage 1978). However, for exposure ages below 100 Myr the small indigenous K content in iron meteorites (typically 10 ppm) significantly increases the uncertainty of the method.

In contrast to iron meteorites where i) the main target elements (Fe, Ni) are abundant, ii) indigenous K contents are negligible, and iii) exposure ages are long (up to 1100 Myr, *e.g.*, Voshage 1967), the OCs and CCs have i) much lower Fe abundance (19−31 wt.%), ii) much higher indigenous K content (~300−900 ppm, **Table 1** and **2**), and iii) significantly shorter exposure times to galactic cosmic rays (typically between 0.1 to 80 Myr; Marti and Graf 1992). Therefore, the influence of cosmogenic K in stony meteorites will be much less evident (if even resolvable). Indeed, a previous survey study on K isotopes show that there is no K isotope variation observed among stony meteorites due to cosmogenic effect within their ~0.5 ‰ best analytical uncertainties (Humayun and Clayton, 1995).

In addition to spallation by primary particles, secondary neutron capture reactions are another possible way to change the isotopic compositions of K in meteorites. Secondary thermal neutrons are also products of cosmic-ray spallation (*e.g.*, on target element Fe). The neutron capture reaction $^{40}$Ca(n, $\gamma$)$^{41}$Ca would generate radioactive $^{41}$Ca, which will eventually decay to $^{41}$K (half-life: 0.1 Myr). This decay is almost instantaneous compared to the long irradiation durations for most of the chondrites (up to 80 Myrs; Graf and Marti 1995). This neutron capture effect would be more pronounced in high Ca/K samples, such as some CCs that contain abundant calcium aluminum-rich inclusions (CAIs). The previous survey study on K isotopes did not observe any K isotope variation among individual CCs and OCs regardless their Ca/K ratios within the ~0.5 ‰ best analytical uncertainties (Humayun and Clayton, 1995). Nevertheless, with the improved analytical precision (~0.05 ‰) in this study, we have observed resolvable variations of $\delta^{41}$K among individual CCs and OCs (see **Figure 2**), and it could be that some of the variation is due to cosmic-ray radiation effects through either direct spallation or secondary neutron-capture reaction.

The Fe/K and Ca/K ratios in different chemical groups of chondrites vary, but within the same chemical groups the Fe/K and Ca/K ratios do not vary significantly (Lodders and Fegley, 1998). Hence, the duration of the cosmic ray irradiation would be the main controlling factor for any variations of $\delta^{41}$K among meteorites from the same chemical group. For example, the cosmic-ray exposure ages of OCs vary from less than 1 Myr, to longer than 80 Myr, with peaks at 6-10 Myr (H), 15 Myr (LL) and 40 Myr (L) (Graf and Marti, 1995). Among CCs, the CI and CM chondrites have the shortest cosmic-ray exposure ages (typically < 2 Myr) compared to CV, CK and CO chondrites (averages:

13 ±10, 23 ±14, and 22 ± 18 Myr, respectively; Eugster et al. 2006). As shown in **Figure 2,** there is no obvious correlation between the average cosmic-ray exposure ages of each group and their K isotopes compositions. For individual meteorites from each group, we plot their K isotopes versus cosmic-ray exposure ages in **Supplementary Figure S1**. Again, we observe no obvious trend of K isotope compositions of individual meteorites versus their cosmic-ray exposure ages. Although cosmogenic spallation or secondary neutron capture reactions remain a possible source of variations in the K isotope compositions of chondrites, such effects cannot be resolved within the current best analytical precisions.

**Supplementary Figure Captions:**

**Supplementary Figure S1**. The bulk K isotopic compositions of carbonaceous chondrites and ordinary chondrites vs. cosmic-ray exposure ages (A) or vs. cosmic-ray exposure ages $(T_e)*[Fe]*[Ca]/[K]$ (B). Exposure age estimates (Hintenberger et al., 1964; Mazor et al., 1970; Eugster, 1988; Herpers et al., 1989; Nishiizumi et al., 1993; Graf and Marti, 1994; Graf and Marti, 1995; Lavielle et al., 1997; Herzog et al., 1997; Scherer and Schultz, 2000; Graf et al., 2001; Leya et al., 2001; Dalcher et al., 2013) are based on cosmogenic noble gas nuclide ($T_3$, $T_{21}$ and $T_{38}$) and $^{36}Cl/^{36}Ar$ methods. A 15 % minimum relative error was assigned to $T_3$ and $T_{21}$ values, following Eugster (1988).

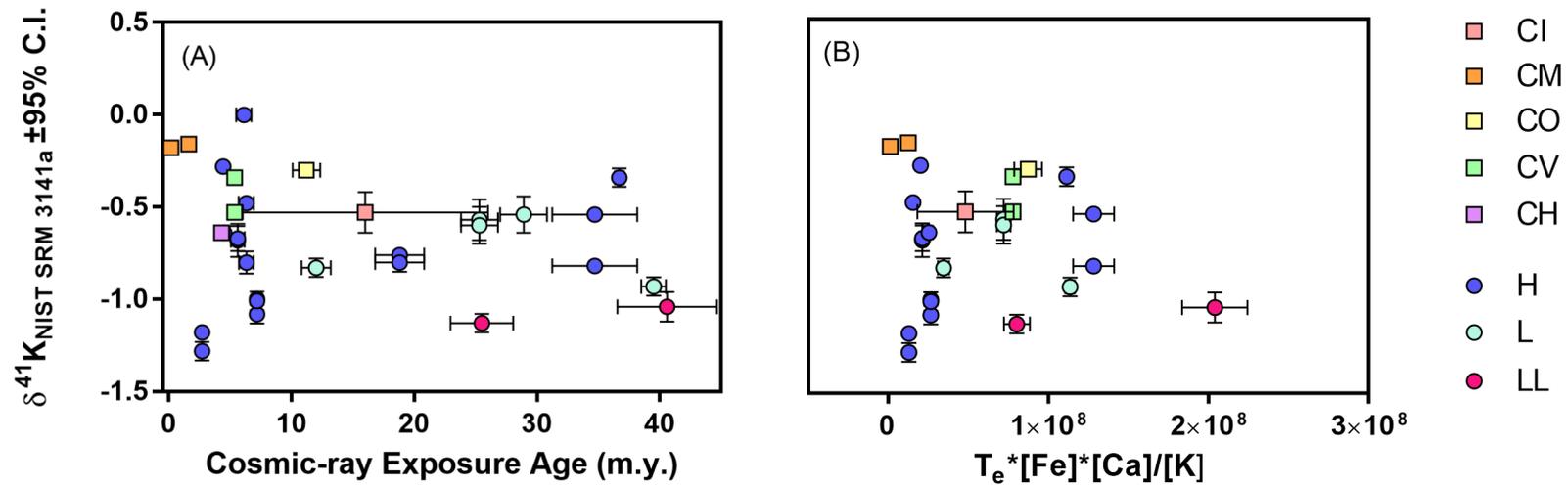

**Supplementary Figure S1**